\documentclass[10pt,twocolumn,twoside]{IEEEtran}

\usepackage{graphicx}
\usepackage{cite}
\usepackage{amsmath}
\usepackage{amssymb,amsthm}
\usepackage{color}
\usepackage[figuresright]{rotating}
\usepackage{subcaption}
\usepackage{graphics}
\usepackage[hidelinks]{hyperref}
\usepackage{fontenc}

\newtheorem{theorem}{Theorem}[section]
\newtheorem{corollary}[theorem]{Corollary}
\newtheorem{lemma}[theorem]{Lemma}

\newtheorem{proposition}[theorem]{Proposition}

\newtheorem{assumption}[theorem]{Assumption}
\newtheorem{definition}[theorem]{Definition}

\newcommand{\V}{\mathcal V}
\newcommand{\A}{\mathcal A}
\newcommand{\Lc}{\mathbf{L}}
\newcommand{\E}{\mathcal E}

\newcommand{\Econ}{\mathcal E_{con}}
\newcommand{\Gcon}{\mathcal G_{con}}
\newcommand{\Vcon}{\mathcal V_{con}}
\newcommand{\Lcon}{\Lc_{con}}
\newcommand{\Lsub}{\Lc_{sub}}

\newcommand{\G}{\mathcal G}
\newcommand{\W}{w}
\newcommand{\N}{\mathcal N}
\newcommand{\Pc}{\mathbf{P}}
\newcommand{\Ab}{\mathbf{A}}
\newcommand{\Ib}{\mathbf{I}}
\newcommand{\Bb}{\mathbf{B}}
\newcommand{\Fb}{\mathbf{F}}

\newcommand{\Hb}{\mathbf{H}}
\newcommand{\Mb}{\mathbf{M}}
\newcommand{\SSigma}{\mathbf{\Sigma}}
\newcommand{\Sb}{\mathbf{S}}
\newcommand{\ssb}{\mathbf{s}}
\newcommand{\Ub}{\mathbf{U}}
\newcommand{\At}{\tilde{\mathbf{A}}}
\newcommand{\xtv}{\tilde{\textbf{x}}}
\newcommand{\vv}{\mathbf{v}}
\newcommand{\xv}{\mathbf{x}}
\newcommand{\wb}{\mathbf{w}}
\newcommand{\Wb}{\mathbf{W}}
\newcommand{\ub}{\mathbf{u}}
\newcommand{\pb}{\mathbf{p}}
\newcommand{\mb}{\mathbf{m}}

\newcommand{\zero}{\mathbf{0}}
\newcommand{\Cis}{C_i^{(2)}}
\newcommand{\Cif}{C_i^{(1)}}

\newcommand{\rb}{\mathbf{r}}

\newcommand{\tp}{^{T}}
\newcommand{\bfo}{\textbf{1}}
\newcommand{\var}[1]{\textbf{var}\left[#1\right]}
\newcommand{\expec}[1]{\textbf{E}\left[#1\right]}
\newcommand{\tr}[1]{\textbf{tr}\left(#1\right)}
\newcommand{\Abar}{\overline{\mathbf{A}}}
\newcommand{\Mcon}{{\widehat{\mathbf{M}}}}
\newcommand{\Mc}{\mathcal{M}}
\newcommand{\Rb}{\mathbf{R}}
\newcommand{\Erdos}{Erd\H{o}s }
\newcommand{\Renyi}{R\'{e}nyi }

\newcommand{\yhy}[1]{{#1}}

\begin{document}

\title{Diffusion and Consensus in a Weakly Coupled Network of Networks}
 \author{Yuhao Yi, Anirban Das, Stacy Patterson, Bassam Bamieh, and Zhongzhi Zhang
\thanks{Yuhao Yi, Anirban Das, and Stacy Patterson are with the Department of Computer Science, Rensselaer Polytechnic Institute, Troy, New York, 12180 USA. {\tt\small yiy3@rpi.edu, dasa2@rpi.edu, sep@cs.rpi.edu}}
 \thanks{Bassam Bamieh is with the Department of Mechanical Engineering, UC Santa Barbara, Santa Barbara, CA 93106 USA. {\tt\small bamieh@ucsb.edu}}
\thanks{Zhongzhi Zhang is with the Shanghai Key Laboratory of Intelligent Information Processing, School of Computer Science, Fudan University, Shanghai, 200433, China. {\tt\small zhangzz@fudan.edu.cn} }
 }
\markboth{}%
{Yi \MakeLowercase{\textit{et al.}}: Diffusion and Consensus in a Weakly Coupled Network of Networks}
%



\IEEEtitleabstractindextext{%
\begin{abstract}
We study diffusion and consensus dynamics in a Network of Networks model. In this model, there is a collection of sub-networks, connected to one another using a small number of links. We consider a setting where the links between networks have small weights, or are used less frequently than links within each sub-network. 
Using spectral perturbation theory, we analyze the diffusion rate and convergence rate of the investigated systems. Our analysis shows that the first order approximation of the diffusion and convergence rates is independent of the topologies of the individual graphs; the rates depend only on the number of nodes in each graph and the topology of the connecting edges. The second order analysis shows a relationship between the diffusion and convergence rates and the information centrality of the connecting nodes within each sub-network. We further highlight these theoretical results through numerical examples. 
\end{abstract}

}

\maketitle

\IEEEdisplaynontitleabstractindextext

%
\IEEEpeerreviewmaketitle

\section{Introduction}

\IEEEPARstart{D}{iffusion} and consensus dynamics play a fundamental role in the coordination of many complex networks, from networks of autonomous vehicles~\cite{ren2008distributed}, to power grids~\cite{6345156}, to social networks~\cite{hegselmann2002opinion}, and beyond.
As such, significant research effort has been devoted to development of analytical characterizations of the performance of diffusion processes and consensus algorithms based on the network topology and the node interactions.  

The vast majority of this work has considered a single, isolated network model.   However, many complex networks can be more accurately represented by a set of interacting networks. For example, in vehicular ad-hoc networks, the network topology often consists of clusters of sub-networks,  made up of co-located vehicles, that periodically communicate with one another~\cite{LW07}.   Another example can be found in social networks, where people are often clustered into communities; interaction within communities is frequent, and interaction across communities less so. These examples motivate the \emph{Network of Networks} (NoN) model, where multiple individual networks, or \emph{subgraphs},  are connected using a few links to form a connected composite graph.

%

We analyze the diffusion rate of an NoN and the convergence rate of consensus algorithms in an NoN using spectral perturbation theory-based methods. 
For the diffusion process in an NoN, we assume the edges between subgraphs has small weights. To formulate this setting, we study a system in which all weights between subgraphs are multiplied with a small parameter $\epsilon$. This setting captures diffusion processes in many complex network systems, for example,  the social networks with weak inter-community links.
In a consensus network,
we consider a setting where the links between subgraphs may be costly to use, and so they are used sparingly in the consensus algorithm. We model this setting using a stochastic system where links that connect subgraphs are active in each iteration with some small probability $p$. This setting applies to architectures like vehicle networks and the Internet of Things, where nearby nodes can communicate using free local communication, e.g., Bluetooth, but where distant nodes must communicate using potentially costly cellular or satellite communication. 

We show that the diffusion rate is directly related to the convergence rate of the expected system of the stochastic consensus network. Our results show that up to first order in $\epsilon$,  the diffusion rate depends on the generalized Laplacian matrix of the connecting graph, which is determined by the number of nodes in each subgraph and the topology of the interconnecting links. The rate does not depend on the topologies of the individual subgraphs nor on which nodes are used to connect the subgraphs to one another. The second order perturbation analysis, however, shows that choosing nodes with largest information centrality~\cite{SZ89} as bridge node maximizes the diffusion rate upto second order in $\epsilon$. We also study the mean square convergence rate of the consensus network, which leads to similar results. 
In addition, we conduct experiments to show that our analysis 
Numerical results show that this analysis accurately captures the behavior of the studied dynamics for small values of $\epsilon$ or $p$. 




\subsubsection*{Related work} 
Several previous papers have studied the diffusion process in various NoN models. \cite{gomez2013diffusion} provides an upper bound for the diffusion rate of a NoN where each layer of subgraph has the same number of nodes and the inter-network links between any two adjacent layers of networks are restricted to be the same one to one map.  \cite{sole2013spectral} studies the same model as \cite{gomez2013diffusion}  using perturbation theory. \cite{tejedor2018diffusion} studies optimal weights for inter-layer links in the case where intra-layer network may be directed. \cite{Cencetti_2019} also studies same model as \cite{gomez2013diffusion} and derives relationships between $\lambda_2$ of the supra-Laplacian and topological properties of the subgraphs. We note that all these works are based on the homogeneous one to one inter-layer connection assumption made in \cite{gomez2013diffusion}. In addition, \cite{SHAO20175184} studies diffusion in Cartesian product of graphs as a model of NoN and gives some analysis based on numerical experiments.

As for the discrete-time consensus dynamics, there has been a significant amount of work devoted to the analysis of distributed consensus algorithms in time-varying networks and stochastic networks, e.g.,~\cite{1205192,1440896,7963624,1393134,zhou2009convergence,abaid2012consensus}. In this work, we employ a model similar to that studied in 
~\cite{Kar08,fagnani2009average,PBE10}, which all study the convergence rate of the mean-square deviation from consensus in a stochastic network.
\cite{Kar08} presents bounds based on the spectrum of the expected weight matrix, whereas~\cite{fagnani2009average} and \cite{PBE10} give analytical expressions
for the convergence rate itself.

None of these previous works considered an NoN model. The NoN consensus model was introduced in \cite{gao2011robustness} and \cite{peixoto2012evolution}, where they measure network performance by analyzing its robustness against random node failures.  More recent work~\cite{7963622} considers an NoN model with noisy consensus dynamics and proposes methods to identify the optimal interconnection topology.  And, in~\cite{7525499}, the authors consider a similar NoN model, but with slightly different dynamics. They show that interconnection between the nodes of subgraphs with the highest degree maximizes the robustness of the NoN. While these works focus on robustness of an NoN, our work in contrast, focuses on the rate at which nodes reach consensus, and in particular, how this rate relates to the topologies of the interconnecting network and the subgraphs.

We note that a preliminary version of this work appeared in~\cite{8619565}. This conference paper presented first-order perturbation analysis only. Further, this analysis was restricted to consensus algorithms, In this paper, we study both diffusion and consensus dynamics, and more significantly, we include second-order perturbation analysis. This second-order analysis provides more insight into the role of the connecting nodes within each subgraph in determining the diffusion and convergence rate.

\subsubsection*{Outline} The rest of the paper is organized as follows. Section~\ref{modelandformulation.sec} describes our system model and the problem formulation, and it gives background on spectral perturbation analysis.
In Section~\ref{expected.sec}, we present analysis of the diffusion and convergence rate in an  NoN, including its first- and second-order behaviors. 
In Section~\ref{msanalysis.eq}, we present our analysis of the mean square convergence of consensus algorithms for a special case of stochastic dynamics. 
 Section~\ref{results.sec} gives numerical evaluations that highlight  key results of our theoretical analysis, followed by the conclusion in Section~\ref{conclusion.sec}.

\section{System Model} \label{modelandformulation.sec}

\subsection{Diffusion Dynamics in Network of Networks} \label{model.sec}
We consider a system of $D$ disjoint graphs ${\G_i = (\mathcal{V}_i, \mathcal{E}_i,\W_i)}$, $i = 1, \ldots, D$. Each graph $\G_i$ is weighted, undirected, and connected. 
We call these graphs the \emph{subgraphs} of the NoN.
The set $\V_i$ denotes the node set of $\G_i$, with $\lvert \V_i \rvert = N_i$, and $\E_i$ is the set of links.   
An edge between node $r \in \V_i$ and $s \in \V_i $  is denoted by $e(r,s)$, and $\N_i(j)$ denotes the neighbor set of node $j$ in subgraph $\G_i$.  
The function $\W_i : \E_i \mapsto \mathbb{R}^+$ defines a non-negative  weight $\W_i(r,s)$ for each edge $e(r,s) \in \E_i$.
Let $\Lc_i$ be the weighted Laplacian matrix of subgraph $\G_i$, defined as
\begin{align*}
\Lc_i(r,s) &= \left\{ \begin{array}{ll} 
\sum_{k \in \N_i(r)} \W_i(r,s) & \text{~for~}r=s \\
-\W_i(r,s) & \text{otherwise.}
\end{array}  \right.
\end{align*}
Further, we define $\Lsub$ to be the $N \times N$ block diagonal matrix with blocks $\Lc_i$, $i = 1 \ldots D$.

We construct an NoN by connecting the $D$  subgraphs with a small number of edges. 
The set $\V$ is the  NoN vertex set, ${\V = \bigcup_{i=1}^D \V_i}$, with $|\V| = N$. Without loss of generality, we identify the nodes in $\V$ as $1, 2, \ldots, N$.
The NoN edge set $\E$ consists of all edges in $\E_1 \cup \ldots \cup \E_D$, as well as a set of undirected \emph{connecting edges} $\Econ =  \{e(r,s)~|~r\in \G_i, s \in \G_j, i \neq j\}$.  
We call the nodes $i \in V$ that are adjacent to some edge in $\Econ$ \emph{connecting nodes}, and we denote the set of connecting nodes by $\Vcon$. 
We assume that there is only one connecting node $s_i$ in each subgraph $\G_i$.
The \emph{connecting graph}  is defined as as $\Gcon = (\V, \Econ, \W_{con})$, where $\W_{con}: \Econ \mapsto \mathbb{R}^+$ is a function that defines a non-negative weight $w_{con}(r,s)$ for each edge $e(r,s) \in \Econ$.
The weighted Laplacian matrix of the connecting graph is denoted by an $N\times N$ matrix $\Lcon$. 
For a matrix $\mathbf{Q}$, we use the symbol $\widehat{\mathbf{Q}}$ to denote the principle submatrix of $\mathbf{Q}$ whose rows and columns correspond to vertices in $\Vcon$.
For example, $\widehat{\Lc}_{con}$ is the $D \times D$ weighted Laplacian of the graph $\widehat{\G}_{con} = (\Vcon, \Econ, \W_{con})$. 

With these definitions, the NoN is thus formally defined as $\G = (\V, \E, \W)$, where $\W(r,s) = \W_i(r,s)$ for $r,s \in \V_i$ and $\W(r,s) = w_{con}(r,s)$ for $r\in \V_i, s\in \V_j, i\neq j$. 
We further define the \emph{strength} of a node $r$ as $\Delta_r = \sum_{s\in \N(r)} w(r,s)$, where $\N(r)$ denotes the neighbor set of node $r$ in graph $\G$.

We study diffusion dynamics in this NoN where there is weak coupling between subgraphs. This weak coupling is enforced both by limiting the number of connecting nodes in each subgraph to one and by selecting a small inter-subgraph diffusion coefficient. 
For each subgraph $\G_i$,  every node $r \in \V_i$ has a scalar-valued state denoted by $x_{r}$.  
The node dynamics are:
\begin{align*}
\dot{x}_r =\!\!\!\! \sum_{s \in \N_i(r)}\!\!\!\! w(r,s) (x_s - x_{r})  + \epsilon\!\! \sum_{e(r,u) \in \Econ} \!\!\!\!\!\!\!\!w(r,s)(x_u - x_r),
\end{align*}
where $\epsilon$ is the diffusion coefficient between subgraphs. 
Let $\xv_i$ denote the vector of node states for graph $\G_i$, and let $\xv$ denote the states of all nodes in the system,
i.e., $\xv = [\xv_1\tp~\xv_2\tp~\ldots~\xv_D\tp]\tp$.
The dynamics of the entire NoN can then be written as:
\begin{align}
\label{contDynamics.eqn}
\dot{\xv} = - (\Lsub + \epsilon \Lcon) \xv.
\end{align}
The matrix $\Lc = \Lsub + \epsilon \Lcon$ is called the \emph{supra-Laplacian} of the NoN. 

We investigate  the smallest non-zero eigenvalue of the Laplacian matrix $\Lc$, which decides the rate of diffusion in (\ref{contDynamics.eqn}).  It is also called the spectral gap of $\Lc$.
\begin{definition}
\label{specGap.def}
The spectral gap of $\Lc = \Lsub + \epsilon \Lcon$ is defined as the smallest non-zero eigenvalue of  $\Lc$, denoted as $\alpha(\Lc)$.
\end{definition}
The spectral gap determines the slowest speed that the diffusion process (\ref{contDynamics.eqn}) converges to its steady state from any initial state and therefore is also referred to as the \emph{diffusion rate}. Since $\Lc$ is positive semi-definite and has eigenvalue zero with multiplicity $1$ for any connected graph $\G$, we know that $\alpha(\Lc)>0$. 
In particular, we study how the spectral gap $\alpha(\Lc)$ is related to matrix $ \Lsub$ and matrix $\Lcon$. We recall that $\Lcon$ is decided by the set of connecting nodes $\V_{con}$ and the structure of the connecting graph, characterized by $\widehat{\Lc}_{con}$. 
Further, we show how are analysis can be used to select connecting nodes within the subgraphs that maximize the spectral gap.



\subsection{Connection to Consensus in Stochastic Networks}
\label{model2.subsection}


%


There is a close relationship between $\alpha(\Lc)$ the convergence rate of discrete-time consensus dynamics in stochastic networks.
 Through this relationship, we identify an alternate interpretation of $\alpha(\Lc)$.
We consider a consensus network where links within each subgraph are always active, e.g., due to the proximity of agents within the subgraph to one another.
Since subgraphs may be separated spatially, communication between subgraphs may be infrequent and/or lossy.
 We model this by activating the connecting edges in $\Econ$  each time step $\ell$ with some small probability $p$.
 One can define the dynamics as a consensus network with stochastic communication links. 
 For a node $r\in \V_i$ 
 \begin{align*}
x_r(\ell+1)&= x_r(\ell) - \sum\limits_{v \in \mathcal{N}_i(r) }w(r,v)\big(x_r(\ell) - x_v(\ell)\big) \\
&\quad \quad - \beta\sum_{e(r,s) \in \Econ} \delta_{rs}(\ell) w(r,s) (x_r(\ell) -x_s(\ell)).
\end{align*}
We assume that for all $r\in \G, r\in \G_i, \sum_{v\in \mathcal{N}_i(r)}{w(r,v)} < 1$. In addition, we assume $\beta \leq \frac{1}{2\Delta}$, where $\Delta = \max(\Delta_i)$ is the maximal node strength of $\G$.
\begin{align*}
\delta_{rs}(\ell) =
\begin{cases} 1 & \text{ with probability } p \\ 
0 & \text{ with probability } 1-p
\end{cases}
\end{align*}
where $\delta_{rs}(\ell)$ are Bernoulli random variables that are not necessarily mutually independent. We note that all $\delta_{rs}(\ell)$ are independent of $\mathbf{x}(\ell)$.

\subsubsection{Convergence Rate of Expected System}

Let  $\mathbf{A}$ be the block diagonal matrix $\mathbf{A} = \Ib - \Lsub$.
We also define an $N \times N$ matrix ${\Bb_{rs} = \beta \cdot w(r,s)\cdot  \textbf{b}_{rs} \textbf{b}_{rs}\tp}$,
where $\textbf{b}_{rs}$ is a binary $N$-vector with the $r^{th}$ element equal to 1, the $s^{th}$ element equal to -1,  and the remaining elements equal to 0. 
The dynamics of the stochastic NoN can then be written as
\begin{align} \label{eq:3}
\mathbf{x}(\ell+1) = \mathbf{A}\mathbf{x}(\ell) - \sum_{e(r,s)\in \Econ}\delta_{rs}(\ell)\Bb_{rs}\mathbf{x}(\ell).
\end{align}

We further let $\bar{\mathbf{x}}(\ell)=\expec{\mathbf{x}(\ell)}$ and $\Bb = \sum_{e(r,s)\in\Econ}\Bb_{rs}$. By taking expectation of both sides of (\ref{eq:3}), we obtain
\begin{align}
\label{eq:4}
\bar{\mathbf{x}}(\ell+1) = \Abar \bar{\mathbf{x}}(\ell)\,,
\end{align}
where $\Abar =  \mathbf{A} -  p\Bb$ is the \emph{expected weight matrix}. The equality follows from the fact that $\delta_{rs}(\ell)$ is independent of $\mathbf{x}(\ell)$.

\begin{definition} 
\label{rhoEss.def}
The \emph{convergence rate of the expected system} of (\ref{eq:4}), denoted $\rho_{ess}(\Abar)$, is defined as the second largest eigenvalue of $\Abar$, also called the \emph{essential spectral radius} of $\Abar$.
\end{definition}
Given the condition $\sum_{v\in \mathcal{N}_i(r)}{w(r,v)} < 1$, the matrix $\Ab_i:= \Ib -  \Lc_i$ has $1$ as a simple eigenvalue with eigenvector $\bfo$ for all subgraph $\G_i$, then matrix $\Ab$ has eigenvalue $1$ with multiplicity $D$. If $\Gcon$ is connected, the matrix $\Abar$ has eigenvalue $1$ with multiplicity $1$, and its corresponding eigenvector is $\bfo$.  Under the assumption $\beta \leq \frac{1}{2\Delta}$, the convergence rate of the expected system (\ref{eq:4}) is characterized by the second largest eigenvalue of $\Abar$~\cite{XIAO200465}. 

Next, noting that $\Ab-p\Bb = \mathbf{I} - (\Lc_{sub}+p\beta\Lc_{con})$,  we state a simple relationship between $\alpha(\Lc)$ and $\rho_{ess}(\Abar)$.
\begin{proposition}
\label{difftoCons.prop}
The spectral gap $\alpha(\Lc)$, where $\Lc = \Lc_{sub}+\epsilon\Lc_{con}$, and the essential spectral radius $\rho_{ess}(\Abar)$, where $\Abar = \Ab-p\Bb$, as given by Definitions~\ref{specGap.def} and~\ref{rhoEss.def}, respectively, satisfy
\begin{align}
\rho_{ess}(\Ab-p\Bb) = 1 - \alpha(\Lc_{sub}+p\beta\Lc_{con})\,.
\end{align}
\end{proposition}

%
\subsubsection{Mean-Square Convergence Rate}
\label{model3.subsubsec}
We also study the \emph{mean square convergence rate} of the stochastic NoN in (\ref{eq:4}). 
Let $\xtv(\ell) = \Pc \xv(\ell)$ be the deviation from average vector,
where $\Pc$ is the projection matrix,  $\Pc = (\Ib_{N} - \frac{1}{N} \bfo \bfo\tp)$.
If $\lim_{t \rightarrow \infty} \expec{ \| \xtv(\ell) \|_2} = 0$, we say the system \emph{converges in mean square}.


We start by investigating the case where all edges in $\G_{con}$ are activated together with some probability $p$ in each time step $t$. We discuss the i.i.d. case in Appendix \ref{extension.sec}.

\begin{assumption}\label{activeTogether.assum}
All edges in $\G_{con}$ are online or offline with probability $p$ and $1-p$ at time step $\ell$, decided by a Bernoulli random variable $\delta(\ell)$.
\end{assumption}

We define the autocorrelation matrix of $\xtv(\ell)$ by $\SSigma(\ell) = \expec{\xtv(\ell) \xtv(\ell)^T}$
and note that $\SSigma(\ell) = \expec{\Pc \xv(\ell) \xv(\ell)\tp \Pc}$.
Using a similar method to that in \cite{patterson2010convergence},  it can be shown that $\SSigma(\ell)$ satisfies the matrix recursion
\begin{align}
\SSigma(\ell+1) &=  (\Pc\bar{\mathbf{A}}\Pc)\SSigma(\ell)(\Pc\bar{\mathbf{A}}\Pc) + \sigma^2 \Bb \SSigma(\ell)\Bb. \label{Mrecur.eq}
\end{align}
where the zero-mean random variable $\mu(\ell)$ is defined as $\mu(\ell) = \delta(\ell) - p$, and $\sigma^2 = \var{\mu(\ell)}$.
The variances  $\mathbb{E}[\tilde{x}_r(\ell)^2]$ are given by the diagonal entries of $\SSigma(\ell)$, and thus we are interested in how they evolve.
We define the matrix-valued operator,
\begin{align}
\A(X) &= (\Pc\bar{\mathbf{A}}\Pc)X(\Pc\bar{\mathbf{A}}\Pc) + \sigma^2 \Bb X \Bb \label{eq:matrixOpDef}
\end{align}
and note that $\SSigma(\ell+1) = \A(\SSigma(\ell))$.
The rate of decay of the entries of $\SSigma(\ell)$ is given by the spectral radius of $\A$, denoted $\rho(\A)$~\cite{patterson2010convergence}.
%
\begin{definition}
The \emph{mean square convergence rate} of the system (\ref{eq:4}), under Assumption~\ref{activeTogether.assum}, is defined as $\rho(\A)$.
\end{definition}


\section{Background on Spectral Perturbation Theory}

Our analytical approach is based on spectral perturbation analysis~\cite{baumgartel85,Ba20}, especially the analysis where repeated eigenvalues are considered~\cite{Ba20}.  Here, we provide a brief overview of this material.

Let $\Mc(\epsilon, X)$ be a symmetric vector-valued (or matrix-valued operator) of a real parameter $\epsilon$ and a variable $X$ of the form
\begin{align}
\Mc(\epsilon, X) = \Mc_0(X)+ \epsilon\Mc_1(X) +\epsilon^2 \Mc_2(X)  \label{eq:secOrderOperator}
\end{align}
and let $(\gamma(\epsilon), W(\epsilon))$  be an eigenvalue-eigenvector (or eigenvalue-eigenmatrix) pair of $\Mc(\epsilon, .)$, as a function of $\epsilon$  
\[
\Mc(\epsilon, W(\epsilon)) = \gamma(\epsilon)W(\epsilon).
\]
According to  spectral perturbation theory, the functions $\gamma$ and $W$ are well-defined and analytic for small  values of ${\epsilon}$. The power series expansion of $\gamma$ is
\begin{align}
\label{perturb.eq}
\gamma(p) = \lambda(\Mc_0) + C^{(1)} \epsilon + C^{(2)} \epsilon^2 + \cdots
\end{align}
where $\lambda(\Mc_0)$ is an eigenvalue of the operator $\Mc_0$.

Let eigenvalue $\lambda(\Mc_0)$ have multiplicity $K$, and let  
$\mathbf{W}_i$, $i=1 \ldots K$,  be $K$ orthonormal eigenvectors (or eigenmatrices)  
of $\Mc_0$ that form a basis for the eigensubspace of $\lambda(\Mc_0)$. 
We form the $K \times K$ matrix $\Fb = [f_{i,j}]$, with each component given by
\begin{align} \label{eq:raleighCoeff}
    f_{ij} = \frac{\langle \mathbf{W}_{i}, \Mc_1(\mathbf{W}_{j})\rangle}{\langle \mathbf{W}_{i}, \mathbf{W}_{i}\rangle}.
\end{align}
When $\Mc$ is a vector-valued operator, the inner product is the standard vector inner product 
(for $\Mc$ a matrix-valued operator, the matrix inner product is $\langle \mathbf{X}, \mathbf{Y}\rangle := \tr{\mathbf{X}^*\mathbf{Y}}$).
Let $\nu_1, \nu_2, \ldots, \nu_{K}$ be the eigenvalues of $\Fb$, with repetition.
Then, the $K$ first-order perturbation constants are $\Cif = \nu_i$, for $i=1 \ldots K$.   

We also study the second order perturbation terms $C^{(2)}$.
 According to \cite{baumgartel85, Ba20}, for an eigenvalue $\lambda(\Mc_0)$ with multiplicity $K > 1$, when $\mathbf{F}$ is diagonal, the second order  terms
 $\Cis$, $i= 1 \ldots K$,  are
\begin{align}
\label{snd_pertb.eq}
\Cis = \sum_{\lambda_m(\Mc_0) \neq \lambda(\Mc_0)} \frac{\langle \Wb_i,\Mc_1(\Wb_m)\rangle^2}{\lambda(\Mc_0)-\lambda_m(\Mc_0)}
\end{align}
where $\Wb_i$ is the $i^{th}$ eigenvector (or eigenmatrix) of $\Mc_0$ with eigenvalue $\lambda$, for $i=1 \ldots K$, and $(\lambda_m(\Mc_0), \Wb_m)$ is an eigenpair of $\Mc_0$ with $\lambda_m(\Mc_0) \neq \lambda(\Mc_0)$.


\section{Analysis} \label{expected.sec}

In this section, we use spectral perturbation analysis to study $\alpha(\Lc)$ and $\rho_{ess}(\Abar)$. 
\subsection{The Spectral Gap in Diffusion Dynamics}
\label{diffAnalysis.subsec}
We first study the convergence of system (\ref{contDynamics.eqn}), assuming the diffusion coefficient $\epsilon$ between subgraphs is small. 
The dynamics in \eqref{contDynamics.eqn} can be expressed using a vector-valued operator of the form given by  (\ref{eq:secOrderOperator}) as $\dot{\xv} = \Mc(\epsilon, \xv)$, where
\begin{align*}
\Mc_o(\xv) &= \Lc_{sub} \xv \\
\Mc_1(\xv) &= \Lc_{con}  \xv \\
\Mc_2(\xv) & = 0.
\end{align*}
We note that $\Lc_{sub}$ is the Laplacian matrix of a graph with $D$ connected components (the subgraphs). Thus, it has an eigenvalue of $0$ with multiplicity $D$. However, when $\widehat{G}_{con}$ is connected, $\Lc$ has an eigenvalue of $0$ with multiplicity $1$. The smallest $D-1$ nonzero eigenvalues of $\Lc$ correspond to the perturbed $0$ eigenvalue of $\Lc_{sub}$. Therefore we study the perturbations to the $0$ eigenvalue 
of $\Lc_{sub}$.


We begin by defining the generalized Laplacian matrix of the connecting graph $\widehat{\G}_{con}$~\cite{rotaru2004dynamic}. 
\begin{definition}
Let $\rb = [N_1~N_2~\ldots~N_D]^T$, and let $\Rb$ be the $D \times D$ diagonal matrix with diagonal entries $\rb$. 
The \emph{generalized Laplacian}of $\widehat{\G}_{con}$ is $\Mcon = \Rb^{-\frac{1}{2}} \widehat{\Lc}_{con} \Rb^{-\frac{1}{2}}$.
\end{definition}
Note that $\Mcon$ is symmetric positive semidefinite. It has an eigenvalue of $0$ with eigenvector $\rb^{1/2}$, and if $\widehat{\G}_{con}$ is connected, its second smallest eigenvalue $\lambda_2(\Mcon)$ is greater than $0$.
We now give a relationship between this eigenvalue and the spectral gap.
\begin{theorem} \label{specGap.thm}
The spectral gap of the matrix $\Lc= \Lsub + \epsilon \Lcon$, up to first order in $\epsilon$,  is
\[
\alpha(\Lc) =  \epsilon \lambda_2(\Mcon)\,,
\]
in which $\lambda_2(\Mcon)$ is the smallest nonzero eigenvalue of $\Mcon$. 
\end{theorem}
\begin{IEEEproof}
We determine the perturbation coefficients  by forming the matrix $\Fb$ in (\ref{eq:raleighCoeff}).
To do so, we must find an orthonormal set of eigenvectors for $D$ zero eigenvalues of $\Lc_{sub}$,  denoted as $\{\vv_1, \ldots, \vv_D\}$.

Let $\ub_1, \ldots, \ub_D$ be  orthonormal eigenvectors of $\Mcon$, and let $\lambda_1(\Mcon) \leq \lambda_2(\Mcon) \leq \ldots \leq \lambda_D(\Mcon)$ be the corresponding eigenvalues.
We define the eigenvectors $\vv_i$, $i=1 \ldots D$, to be ${\vv_i = [ \theta_i^{(1)} \bfo_{N_1}\tp~ \theta_i^{(2)} \bfo_{N_2}\tp~\ldots~\theta_i^{(D)} \bfo_{N_D}\tp]\tp}$, with
\begin{equation} \label{eq:thetaDef2}
\theta_i^{(j)} = \frac{1}{\sqrt{N_j}} u_{ij}
\end{equation}
where $u_{ij}$ denotes the $j^{th}$ component of the eigenvector $\ub_i$. We observe that the eigenvectors $\vv_i$, $i=1 \ldots D$, are orthonormal.

We now find the entries of the $D \times D$ matrix $\Fb$ defined by (\ref{eq:raleighCoeff}). 
For $f_{ij}$, we have
\begin{align}
f_{ij} &=  \langle \vv_i, \Lc_{con} \vv_j \rangle \nonumber \\
&= \ub_j \tp \Rb^{-\frac{1}{2}} \widehat{\Lc}_{con} \Rb^{-\frac{1}{2}}  \ub_i  \nonumber \\
&=  \ub_j \tp \Mcon \ub_i \nonumber \\
&=  \lambda_i(\Mcon) \ub_i^T \ub_j. \label{gijgen.eq}
\end{align}
The equalities follow by the definition of $\vv_i$, $\widehat{\Mb}$, and $\ub_i$.
If $i \neq j$, then because $\ub_i$ and $\ub_j$ are orthonormal, $f_{ij} = 0$. 
Thus $\Fb$ is a diagonal matrix, and its eigenvalues are
\begin{equation}
C_{i}^{(1)} =  \lambda_i(\Mcon),~i = 1 \ldots D. \label{gii.eq}
\end{equation}
This completes the proof.
\end{IEEEproof}

Theorem~\ref{specGap.thm} shows that the diffusion rate, up to first order in $\epsilon$, is decided by an expression that depends on the smallest nonzero eigenvalue of $\Mcon$. We note that $\Mcon$ depends on the topology and edge weights of the connecting graph, as well as the number of vertices in each subgraph. However, $\Mcon$ does not depend on the topology or edge weights of the subgraphs. Further, it does not depend on the choice of connecting node in each subgraph.
An intuition for this result is that the connecting link is a bottleneck in the diffusion process.
The diffusion rate within each graph is much faster than the diffusion rate across the connecting link.
The role of the connecting link is to transfer information between the two graphs, and the amount of information that needs to be exchanged is proportional to the sizes of the graphs. 
It has been shown that $\lambda_i(\Mcon)$ also determines the convergence rate of load balancing diffusion algorithms in heterogeneous systems~\cite{rotaru2004dynamic}. Following this analogy,  we can view just the edges in $\Gcon$  as executing a load balancing algorithm. The role of the connecting graph is to transfer load (i.e., node state) between the  subgraphs, and the load that needs to be transferred out of each subgraph to balance the system is be proportional to the number of nodes in that subgraph.

Then we study the diffusion rate of (\ref{contDynamics.eqn}) upto second order of $\epsilon$. We note that it is decided by the spectral gap of $\Lc$.
\begin{theorem}
\label{sndPertAlpha.thm}
The spectral gap of the matrix $\Lc= \Lsub + \epsilon \Lcon$, up to second order in $\epsilon$, is
\begin{align}
\label{sndPertAlpha.eqn}
\alpha(\Lc) =  \epsilon \lambda_2(\Mcon)-\epsilon^2 ((\lambda_2(\Mcon))^2 (\ub_2^* \widehat{\mathcal{S}} \ub_2)\,,
\end{align}
where $\lambda_2(\Mcon)$ is the smallest nonzero eigenvalue of $\Mcon$, and $\ub_2$ is its corresponding eigenvector.  The $D\times D$ diagonal matrix $\widehat{\mathcal{S}}$ has diagonal entries $\widehat{\mathcal{S}}(k,k):=N_k \cdot \mathbf{L}_k^{\dag}(s_k,s_k)$. $\mathbf{L}_k^{\dag}$ is the Moore-Penrose inverse of $\mathbf{L}_k$, $s_k$ is the connecting node in graph $\G_k$, and $\mathbf{L}_k^{\dag}(s_k,s_k)$ is the diagonal entry of $\mathbf{L}_k$ that corresponds to node $s_k$.
\end{theorem}
\begin{IEEEproof}
In order to study second order perturbation coefficients using (\ref{snd_pertb.eq}), we need to find all $N$ eigenvectors of the matrix $\Lsub$.

We recall that the eigenvectors of $\Lsub$ corresponding to zero eigenvalues are defined as ${\vv_i = [ \theta_i^{(1)} \bfo_{N_1}\tp~ \theta_i^{(2)} \bfo_{N_2}\tp~\ldots~\theta_i^{(D)} \bfo_{N_D}\tp]\tp}$, where $\theta_i^{(1)}$ is defined by (\ref{eq:thetaDef2}), for $i=1\ldots D$.

We define the remaining eigenvectors of $\Lc_{sub}$ as follows. 
Consider the Laplacian matrix $\Lc_i$ for subgraph $i$, and let $\pb_{i_\psi}$, ${\psi=1 \ldots N_i}$, be a set of $\psi$ orthonormal eigenvectors of $\Lc_i$. Since $\G_i$ is connected, its $0$ eigenvalue has multiplicity $1$. We let $\pb_{i_\psi}$, ${\psi=2 \ldots N_i}$, be the eigenvectors associated with nonzero eigenvalues.
Then we define the remaining $\vv_m$, $m=(D+1) \dots N$, to be ${\vv_m = [ \zero_{N_1}\tp~\ldots \zero_{N_{k-1}}\tp~\pb_m\tp~\zero_{N_{k+1}}\tp~\ldots~\zero_{N_{D}}\tp]\tp}$, where $\pb_m \in \{\pb_{i_\psi} : i\in[D] \textrm{ and }\psi\in\{2,\ldots, N_i\}\}$. 

By applying (\ref{snd_pertb.eq}) we attain
\begin{align}
C_{i}^{(2)}  &  = \sum_{\lambda_m(\Lc_{sub})\neq 0}\frac{\vv_i^*\Lc_{con}\vv_m\vv_m^*\Lc_{con}\vv_i}{0-\lambda_m(\Lc_{sub})} \nonumber\\
& = \sum_{\lambda_m(\Lc_{sub})\neq 0}\frac{\widehat{\vv}_i^*\widehat{\Lc}_{con}\widehat{\vv}_m\widehat{\vv}_m^*\widehat{\Lc}_{con}\widehat{\vv}_i}{0-\lambda_m(\Lc_{sub})}\nonumber
\end{align}
We recall that $s_k$ is the vertex index of the connecting node in subgraph $\G_k$. Then
\begin{align}
& C_{i}^{(2)}  = \sum_{k=1}^D\sum_{\substack{m: \\supp(\vv_m)\subset \V_k}}\frac{\widehat{\vv}_i^*\widehat{\Lc}_{con}(p_{m,s_k}^2\mathbf{E}_{k})\widehat{\Lc}_{con}\widehat{\vv}_i}{-\lambda_m(\Lc_k)}\nonumber\\
& =\sum_{k=1}^D\widehat{\vv}_i^*\widehat{\Lc}_{con}\Rb^{-\frac{1}{2}}\left(\sum_{\substack{m: \\supp(\vv_m)\subset \V_k}}\!\!\!\!\!\!\!\!\frac{p_{m,s_k}^2\Rb^{\frac{1}{2}}\mathbf{E}_{k}\Rb^{\frac{1}{2}}}{-\lambda_m(\Lc_k)}\right)\Rb^{-\frac{1}{2}}\widehat{\Lc}_{con}\widehat{\vv}_i \nonumber\\
& = \sum_{k=1}^D\widehat{\vv}_i^*\widehat{\Lc}_{con}\Rb^{-\frac{1}{2}}\left(\sum_{\substack{m: \\supp(\vv_m)\subset \V_k}}\!\!\!\!\!\!\!\!\frac{r_{kk}\cdot p_{m,s_k}^2\mathbf{E}_{k}}{-\lambda_m(\Lc_k)}\right)\Rb^{-\frac{1}{2}}\widehat{\Lc}_{con}\widehat{\vv}_i\,,\nonumber
\end{align}
where $\mathbf{E}_k$ is a $D\times D$ matrix with only one non-zero entry $\mathbf{E}_{k,k} = 1$. $p_{m,s_k}$ is the entry of $\mathbf{p}_{m}$ associated with the connecting node $s_k$.
We can further derive
\begin{align}
C_{i}^{(2)}&= -\ub_i^*\Rb^{-\frac{1}{2}}\widehat{\Lc}_{con}\Rb^{-\frac{1}{2}}\widehat{\mathcal{S}}\Rb^{-\frac{1}{2}}\widehat{\Lc}_{con}\Rb^{-\frac{1}{2}}\ub_i\nonumber\\
& =  -\ub_i^* \Mcon \widehat{\mathcal{S}} \Mcon \ub_i \nonumber\\
& = - (\lambda_i(\Mcon))^2 (\ub_i^* \widehat{\mathcal{S}} \ub_i)
\,,\label{expt2nd.eq}
\end{align}
where the $D\times D$ diagonal matrix $\widehat{\mathcal{S}}$ has its entries $\widehat{\mathcal{S}}(k,k):=r_{kk}\cdot \mathbf{L}_k^{\dag}(s_k,s_k)$.
From (\ref{perturb.eq}) we attain the result given in Theorem \ref{sndPertAlpha.thm}.
\end{IEEEproof}

We further obtain the following corollary for all the eigenvalues of $\Lc$ up to first order and second order in $\epsilon$.
\begin{corollary}
\label{bridgenode.thm}
For any nonzero eigenvalue $\lambda_i(\Lc)$, $i= 2,\dots,D$ in the studied network of networks system (\ref{eq:3}), the first order approximation of $\lambda_i(\Lc)$ is independent of the choices of connecting nodes, the second order approximation of $\lambda_i(\Lc)$ is \yhy{maximized} when each connecting node is chosen as the one with maximum information centrality in each subgraph.
\end{corollary}
\begin{IEEEproof}
From Theorem \ref{specGap.thm} we know that the first order approximation of $\lambda_i(\Lc)$ does not depend on the choice of the connecting nodes.


Then we take into account the second order perturbation terms given by (\ref{expt2nd.eq}). We note that once the structure and the weight function of the connecting graph are fixed, $\lambda_i(\Mcon)$ and $\ub_i$ are determined for all $i$. As long as the choice of connecting nodes is concerned, $C^{(2)}_{i}$ is maximized when $\widehat{\mathcal{S}}$ is minimized in the Loewner order. This is achieved when the diagonal entries $\widehat{\mathcal{S}}(k,k)$ are all minimized simultaneously. This is then achieved when each bridge node is chosen as the node with maximum information centrality~\cite{SZ89} in that subgraph, because $r_{k,k} = N_k$  is the same for any choice in that subgraph.
\end{IEEEproof}

Corollary \ref{bridgenode.thm} shows that the second-order perturbation terms are affected by the choice of connecting node in each subgraph. The second-order approximations of all eigenvalues are maximized simultaneously when each connecting node is chosen as the node with maximum information centrality in the corresponding subgraph.



\subsection{Analytical Examples}
\subsubsection{Analysis for $D=2$}

For an NoN consisting of two subgraphs $\G_1$ and $\G_2$, the backbone graph $\Gcon$ consists of a single edge.
\begin{corollary}
\label{dif2subgraph.thm}
The spectral gap $\alpha(\Lc)$ of an NoN consisting of two subgraphs $\G_1$ and $\G_2$, up to first order in $\epsilon$, is
\begin{align}
\alpha(\Lc)=  \epsilon  \left( \frac{N}{N_1N_2}\right)
\label{alpha2.eq}
\end{align}
\end{corollary}
\begin{IEEEproof}
The generalized Laplacian matrix $\Mcon$ is given by
\[
\Mcon =
\begin{bmatrix}
\frac{1}{N_1} & -\frac{1}{\sqrt{N_1 N_2}}\\
-\frac{1}{\sqrt{N_1 N_2}} & \frac{1}{N_2}
\end{bmatrix}
\]
$\Mcon$ has two eigenvalues, $\lambda_1(\Mcon)=0$ and $\lambda_2(\Mcon)=\frac{1}{N_1}+ \frac{1}{N_2}$. Their corresponding eigenvectors are $\ub_1 \!\!=\!\! \frac{1}{\sqrt{N}} [\sqrt{N_1}\, \!\sqrt{N_2}]^{T}$ and $\ub_2=\frac{1}{\sqrt{N}} [\sqrt{N_2}\, -\sqrt{N_1}]^{T}$. Applying the definition for $F_{i}^{(1)}$ in (\ref{gii.eq}), we obtain (\ref{alpha2.eq}).
\end{IEEEproof}


This theorem shows that the first order approximation of $\alpha(\Lc)$ depends on the number of nodes in each subgraph. The first order approximation does not depend on the structures of the subgraphs or the choice of bridge node within each subgraph, as we have observed in Theorem~\ref{specGap.thm}.

We can also observe from (\ref{alpha2.eq}) that when $N_1=N_2 = \frac{N}{2}$, the first order approximation of $\alpha(\Lc)$ is minimized. In other words, when two subgraphs have the same number of nodes, the system converges rate is smallest. 


\subsubsection{Analysis for $D>2$ with Equally Sized Graphs}

We next consider the case where $N_1 = N_2 = \ldots = N_D = \frac{N}{D}$, i.e., all subgraphs have the same number of nodes.

\begin{corollary} \label{equal.thm}
Consider  a composite system consisting of $D$ subgraphs $\G_1, \ldots, \G_D$, each with $\frac{N}{D}$ nodes, and a backbone graph $\Gcon$.  The spectral gap $\alpha(\Lc)$, up to first order in $\epsilon$, is
\begin{align*}
    \alpha(\Lc) = \epsilon  \left(\frac{D}{N}\right) \lambda_2(\widehat{\Lc}_{con}) \,.
\end{align*}
where  $\lambda_2(\widehat{\Lc})_{con}$ is the second smallest eigenvalue of $\widehat{\Lc}_{con}$. 
\end{corollary}

\begin{IEEEproof}
Given $N_1 = N_2 = \ldots = N_D = \frac{N}{D}$, we attain $\Mcon = \frac{D}{N}\widehat{\Lc}_{con}$. Then we obtain the result in Corollary \ref{equal.thm} by applying Theorem \ref{specGap.thm}.
\end{IEEEproof}

As with the case where $D=2$, up to the first order approximation, the convergence factor is independent of the topology of the subgraphs, and it is independent of the choice of connecting nodes.
The diffusion rate depends on $\lambda_2(\Lcon)$, also called the \emph{algebraic connectivity} of the backbone graph. 
If $\Gcon$ is not connected, then $\lambda_2(\Lcon) = 0$, meaning, as expected, the system does not converge. 
The diffusion rate increases as the algebraic connectivity of $\Gcon$ increases. 

\subsection{Convergence Rate of the Expected Consensus Network}
Next we study the convergence rate of the the expected consensus network~(\ref{eq:4}). By using the analytic results we developed in \ref{diffAnalysis.subsec}, as well as the connection between the spectral gap of $\Lc$ and the essential spectral radius of $\Abar$, we obtain the following corollary.
\begin{corollary} \label{expectedrho.thm}
The essential spectral radius of the expected weight matrix $\Abar$, up to first order in $p$,  is
\[
\rho_{ess}(\Abar) = 1 - p \beta \lambda_2(\Mcon);
\]
the essential spectral radius, upto second order in $p$, is
\[
\rho_{ess}(\Abar) = 1 - p \beta \lambda_2(\Mcon) +p^2 \beta^2 ((\lambda_2(\Mcon))^2 (\ub_2^* \widehat{\mathcal{S}} \ub_2).
\]
\end{corollary}
We omit the proof of Corollary~\ref{expectedrho.thm} because the results follow straightforwardly from Proposition \ref{difftoCons.prop}, Theorem~\ref{specGap.thm}, and Theorem~\ref{sndPertAlpha.thm}.

According to Proposition \ref{difftoCons.prop} and Corollary~\ref{bridgenode.thm}, we conclude that the first order approximation of $\rho_{ess}(\Abar)$ is independent of the choices of connecting nodes; the second order approximation shows that choosing nodes with maximum information centrality as the connecting node in each subgraph leads to the fastest convergence rate for the expected consensus system (\ref{eq:4}).

\section{Analysis of Mean Square Convergence Rate} \label{msanalysis.eq}
We now use spectral perturbation analysis to study the mean square convergence rate of an NoN in which all edges in $\Econ$ are activated together with small probability $p$. 

\subsection{Mean Square Perturbation}



We write the operator in (\ref{eq:matrixOpDef}) as a matrix-valued operator $\A(X,p)$ of both a matrix $X$ and the small probability $p$in the form (\ref{eq:secOrderOperator}), with
\begin{align}
\A_0(X) &= \At X\At  \label{eq:A0} \\
\A_1(X) &= -\Bb X\At  - \At X \Bb  + \Bb X \Bb \label{eq:A1} \\
\A_2(X) &= \Bb X\Bb -\Bb X\Bb = \mathbf{0} \label{eq.A2}\,,
\end{align}
where $\At = \Pc \Ab \Pc$. Recall that $\Ab = \Ib - \Lsub$.
Given the assumption that for all $r\in \G, r\in \G_i, \sum_{v\in \mathcal{N}_i(r)}{w(r,v)} < 1$, then for each subgraph $\G_i$, $\Lc_i$ has a single $0$ eigenvalue. Then the matrix $\Lsub$ has eigenvalue $0$ with multiplicity $D$,
it follows that $\At = \Pc - \mathbf{L}$ has eigenvalue $1$ with multiplicity $D-1$.  Therefore, the operator $\A_0$ has an eigenvalue of $1$ with multiplicity $(D-1)^2$. 
When the system is perturbed by $p \A_1$, these $1$ eigenvalues are  perturbed. The perturbed eigenvalue with largest magnitude is $\rho(\A)$.


For any pair of eigenvectors $\wb_i$ and $\wb_j$ of the matrix $\At$, $\Wb_{ij}:=\wb_i\wb_j^*$ is an eigenmatrix of $\A_0$ with eigenvalue $\lambda_{ij}(\A_0)=\lambda_i(\At)\lambda_j(\At)$.   Because $\At = \Pc - \mathbf{L}$ is symmetric, its left and right eigenvectors satisfy $\wb_i^*\wb_i=1$ for $i\in[N]$ and $\wb_i^*\wb_j=0$ for any $i,j\in[N]$, $i\neq j$.

\begin{lemma} \label{rhoA.lem} 
Let $\G = (\V,\E)$ be an NoN with the dynamics as defined in (\ref{eq:3}). There exists a set of vectors $\{\wb_i :i = 2,\ldots,D\}$ and an induced set of matrices $\{{\mathbf{W}_{ij}}= \wb_i\wb_j^*: i,j\in \{2,\ldots,D\}\}$ such that 
\begin{align}
 \A_0({\mathbf{W}_{ij}}) = {\mathbf{W}_{ij}}, \qquad&\forall i,j \in \{2,\ldots,D\}\,,\label{prop1}\\
 \wb_i^*\wb_i = 1, \qquad &\forall i,  \{2,\ldots,D\}\,,\label{prop2}\\
 \wb_i^*\wb_j = 0, \qquad &\forall i,j\in  \{2,\ldots,D\}, i \neq j\,\label{prop3}\\
  \wb_i^*\bfo = 0, \qquad &\forall i\in  \{2,\ldots,D\}\,,\\
 \wb_i^*\Bb\wb_j = 0, \qquad &\forall i,j\in  \{2,\ldots,D\}, i \neq j \label{prop5}
\end{align}
The mean square convergence rate of system \yhy{(\ref{eq:3})} satisfying Assumption \ref{activeTogether.assum}, up to first order in $p$, is
\begin{align}
\rho(\A) = \max_{ij} \left(1 + p f^{(1)}_{ij} \right)\,,
\end{align}
in which
\begin{align}
    f^{(1)}_{ij} = -  \wb_i^* \Bb \wb_i - \wb_j^* \Bb \wb_j + 
    \left( \wb_i^* \Bb \wb_i \right) \left( \wb_j^* \Bb \wb_j   \right)\,. \label{fii.eq}
\end{align}
\end{lemma}
\begin{IEEEproof}
Let  $\textbf{M}_{ij} = \mb_i \mb_j^*$, $i,j\in  \{2\ldots D\}$ be any set of (mutual) orthonormal eigenmatrices of $\A_0$  associated with eigenvalue $1$.
The vectors  $\mb_i$, $i=2 \ldots D$ are eigenvectors of $\At$ such that $\At \mb_i = \mb_i$; further, they are mutually orthonormal and are all orthogonal to the vector $\bfo$. 

We define a matrix $\Hb$ whose entries are defined as $h_{ij} = \mb_i^*\Bb\mb_j$. Let $\Ub$ be the matrix whose columns are $\mb_i$, $i\in  \{2\ldots D\}$. Then it is clear that $\Hb = \Ub^*\Bb\Ub$. Let $\Hb = \Sb \mathbf{\Lambda} \Sb^*$ be the spectral decomposition of $\Hb$. $\Sb$ is an unitary matrix, $\ssb_i$ is the $i$th column of $\Sb$.  Therefore $\Bb = \Ub\Sb\mathbf{\Lambda}\Sb^*\Ub^*$. We define $\wb_i := \Ub\ssb_i$, for all $i\in  \{2\ldots D\}$. It is easy to verify that the vectors in $\{\wb_i :i = 2,\ldots,D\}$ satisfy the properties (\ref{prop1})-(\ref{prop5}) stated in the lemma.
We note that by (\ref{prop2}) and (\ref{prop3}), ${{\langle \mathbf{W}_{ij}, \mathbf{W}_{ij}\rangle} = 1}$ for all $i,j\in  \{2\ldots D\}$; ${{\langle \mathbf{W}_{ij}, \mathbf{W}_{pq}\rangle} = 0}$ for all $i\neq p$ or $j\neq q$. Therefore, we consider the entries of the $(D-1)^2\times (D-1)^2$ matrix $\Fb$:
\begin{align}
f_{ij,pq} &= \langle \wb_i\wb_j^*, \A_1(\wb_p \wb_q^*)\rangle  \nonumber \\
&= \tr{\wb_j\wb_i^*\left(-\Bb \wb_p \wb_q^* \At - \At \wb_p \wb_q^* \Bb + \Bb \wb_p \wb_q^* \Bb\right)} \nonumber \\
&= -\tr{\wb_j \wb_i^*\Bb \wb_p \wb_q^*} - \tr{\wb_j \wb_i^*\wb_p \wb_q^* \Bb}  \nonumber  \\
&~~~~~+ \tr{\wb_j \wb_i^* \Bb \wb_p \wb_q^* \Bb} \label{f3.eq} 
\end{align}
where the last equality holds since ${\At \wb_p = \wb_p}$ and similarly, ${\wb_q^* \At = \wb_q^*}$.
the expression can further be written as
\begin{align*}
f_{ij,pq} =&  - \wb_i^* \Bb \wb_p \wb_q^* \wb_j  - \wb_i^* \wb_p \wb_q^* \Bb \wb_j \nonumber\\
& +\left( \wb_i^* \Bb \wb_p \right) \left( \wb_j^* \Bb \wb_q   \right). 
\end{align*}

If $i=p$ and $j=q$, then noting that $\wb_i^* \wb_p = 1$ and $\wb_j^*\wb_q=1$, it follows that
\begin{align*}
    f^{(1)}_{ij} := f_{ij,ij} = -  \wb_i^* \Bb \wb_i - \wb_j^* \Bb \wb_j + 
    \left( \wb_i^* \Bb \wb_i \right) \left( \wb_j^* \Bb \wb_j   \right). 
\end{align*}
Furthermore, since $\wb_i^*\Bb\wb_j=0$ for any $i\neq j$, all off diagonal entries are zeros.
\end{IEEEproof}

We next use this lemma to characterize the convergence factor in two classes of NoNs.

\subsection{Analysis for Special Cases}
We give results for the mean square convergence rate for the two cases which we have discussed in Section~\ref{expected.sec}.
\begin{corollary}
\label{2subgraph.thm}
For an NoN  consisting of two subgraphs $\G_1$ and $\G_2$, with the dynamics \yhy{(\ref{eq:3})} satisfying Assumption \ref{activeTogether.assum}, the mean square convergence rate, up to first order in $p$, is
\begin{align}
\rho(\A) = & 1- 2p\beta  \left( \frac{N}{N_1N_2}\right) + p\beta^2 \left( \frac{N}{N_1N_2}\right)^2\,.
\label{rho2.eq}
\end{align} 
\end{corollary}
\begin{IEEEproof}
We define the vector $\wb_2$ as
\[
\wb_2 =
\begin{bmatrix}
\theta^{(1)} \boldsymbol{1}_{N_1}\\
\theta^{(2)} \boldsymbol{1}_{N_2}\,.
\end{bmatrix}
\]
where $\theta^{(1)}= \sqrt{\frac{N_2}{N \cdot N_1}}$ and $\theta^{(2)}= -\sqrt{\frac{N_1}{N \cdot N_2}}$. 
It is easily observed that $\wb_2$ is an eigenvector of $\At$ with eigenvalue 1, and $\wb_2$ is orthogonal to $\bfo$.
 When $D=2$, the matrix $\Fb$  consists of a single element. Applying the definition for $f^{(1)}_{22}$ in (\ref{fii.eq}), we obtain
\begin{align}\label{eq:f11}
    f^{(1)}_{22} &=-2\beta(\theta^{(1)}-\theta^{(2)})^2
+ \beta^2 ( \theta^{(1)}  -  \theta^{(2)})^4\\
& = -2\beta\left(\frac{N}{N_1N_2}\right)
+ \beta^2\left(\frac{N}{N_1N_2} \right)^2\,.
\end{align}
This completes the proof.
\end{IEEEproof}
From (\ref{rho2.eq}) we observe that given $N$, the magnitude of $\rho(\A)$ is maximized when the graphs are of the same size, i.e., $N_1 = N_2$. It is minimized when $N_1=1$, $N_2=N-1$ or $N_2=1$, $N_1=N-1$. This means that the speed of convergence is slower between balanced subgraphs. By comparing (\ref{rho2.eq}) to (\ref{alpha2.eq}) we note that for two subgraphs, both $\rho_{ess}(\Abar)$ and $\rho(\A)$ are determined by the strength (activation probability) of the connecting edge and the number of nodes in both subgraphs.


%
%


\begin{corollary} \label{equal2.thm}
For an NoN consisting of $D$ subgraphs $\G_1, \ldots, \G_D$, each with $\frac{N}{D}$ nodes, with the system dynamics \yhy{(\ref{eq:3})} satisfying Assumption \ref{activeTogether.assum}, the mean square convergence factor, up to first order in $p$, is
\begin{align*}
    \rho(\A) =& 1 -p \left(2 \beta  \left(\frac{D}{N}\right) \lambda_2(\widehat{\Lc}_{con})  {-} \beta^2  \left(\frac{D}{N}\right)^2 (\lambda_2(\widehat{\Lc}_{con}))^2 \right) \,.
\end{align*}
where  $\lambda_2(\Lcon)$ is the second smallest eigenvalue of $\Lcon$. 

%
\end{corollary}

\begin{IEEEproof}
We obtain this result by defining the  $D-1$ eigenvectors of $\At$ with eigenvalue 1 as follows.
Let $\ub_1, \ldots, \ub_{D}$ be an orthonormal set of eigenvectors of the $D \times D$ matrix $\widehat{\Lc}_{con}$ with eigenvalues $0=\lambda_1(\widehat{\Lc}_{con}) \leq  \ldots \leq \lambda_{D}(\widehat{\Lc}_{con})$.  
Let  ${\ub_1 = (1/\sqrt{D}) \bfo}$, and thus $\Lcon \ub_0 = 0$.
The $i^{th}$ eigenvector of $\At$, $i=2 \ldots D$, is  
\[
\wb_i = [ \theta_i^{(1)} \bfo_{N_1}\tp~~ \theta_i^{(2)} \bfo_{N_2}\tp~~~\ldots~~~\theta_i^{(D)} \bfo_{N_D}\tp]\tp
\]
with
\begin{equation} \label{eq:thetaDef}
\theta_i^{(j)} = \frac{1}{\sqrt{N/D}} u_{ij}
\end{equation}
where $u_{ij}$ denotes the $j^{th}$ component of the eigenvector $\ub_i$, $j=1 \ldots D$. 
Therefore, the first perturbation term of the eigenvalue corresponds to eigenmatrix $\Wb_{ij}=\wb_i\wb_j^*$ are obtained:
\begin{align}
\label{fijequal.eq}
f_{ij}^{(1)} = &-\beta\left(\frac{D}{N}\right)\left(\lambda_i(\widehat{\Lc}_{con})+\lambda_j(\widehat{\Lc}_{con})\right) \nonumber\\
&+ \beta^2  \left(\frac{D}{N}\right)^2 \lambda_i(\widehat{\Lc}_{con})\lambda_j(\widehat{\Lc}_{con})\,.
\end{align}

By Lemma~\ref{rhoA.lem} and (\ref{fijequal.eq}), $\rho(A)$ is equal to 
\begin{align}
\rho(\A)&=\max_{i,j \in \{2, \ldots, D\}} 1 -p \left(2 \beta  \left(\frac{D}{N}\right) \left( \lambda_i(\widehat{\Lc}_{con}) + \lambda_j(\widehat{\Lc}_{con}) \right) \right.\nonumber\\
&~~~~~~~~~~~~~~~\left. 
-  \beta^2  \left(\frac{D}{N}\right) ^2  \lambda_i(\widehat{\Lc}_{con})\lambda_j(\widehat{\Lc}_{con}) \right). \label{maxequal.eq}
\end{align}
The maximum node degree of any node $v \in \Vcon$ is $D-1$; thus, the eigenvalues of $\widehat{\Lc}_{con}$ are in the interval ${[0, 2\Delta]}$~\cite{merris1994laplacian}.
Since $\beta < \frac{1}{2\Delta}$, we have $\beta  \lambda_j(\widehat{\Lc}_{con})   \in [0,1)$ for $j=2 \ldots D$. Further we attain that $2\beta(\frac{D}{N})-\beta^2(\frac{D}{N})^2\lambda_j(\widehat{\Lc}_{con})>0$ for $j=2\ldots D$. Thus, the right hand side of expression (\ref{maxequal.eq}) is maximized when $\lambda_i(\widehat{\Lc}_{con})$ is minimized. The same analysis holds for $\lambda_j(\widehat{\Lc}_{con})$. So the right hand side of expression (\ref{maxequal.eq}) is maximized when both $\lambda_i(\widehat{\Lc}_{con})$ and $\lambda_j(\widehat{\Lc}_{con})$ are equal to $\lambda_2(\widehat{\Lc}_{con})$, which proves the theorem.
\end{IEEEproof}
We observe from Corollary~\ref{equal2.thm} and Corollary~\ref{equal.thm} that for subgraphs with the same number of nodes, both $\rho_{ess}(\Abar)$ and $\rho(\A)$ are determined by the algebraic connectivity of the connecting graph as well as the number of nodes in each subgraph.

We note that the second-order perturbation analysis similar to Theorem \ref{bridgenode.thm} can also be applied to the analysis of mean-square convergence rate of (\ref{eq:4}) satisfying Assumption~\ref{activeTogether.assum}. We defer the related discussion to Appendix \ref{appendix2.sec}.


\section{Numerical Results} \label{results.sec}

In this section, we give some numerical examples to support our analytic results. Edges are weighted $1$ in these examples unless otherwise specified.
All experiments were done in MATLAB.

\begin{figure*}[htbp]
	\begin{subfigure}[t]{0.32\linewidth}     
      \centering
      \includegraphics[width=\linewidth]{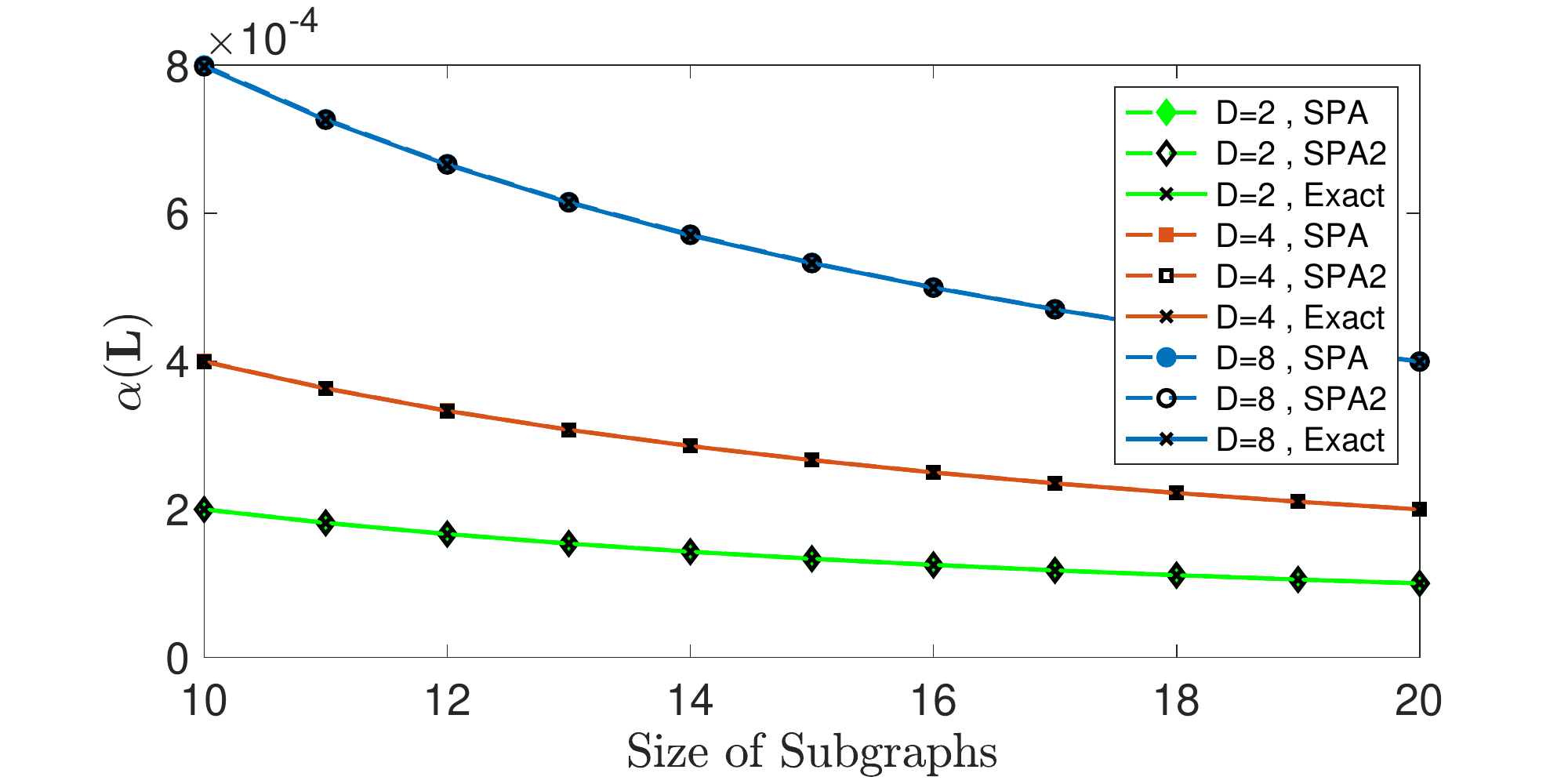}
      \subcaption{$\epsilon = 0.001$}
      \label{GD1_1}
   \end{subfigure}%
   \begin{subfigure}[t]{0.32\linewidth}
      \centering
      \includegraphics[width=\linewidth]{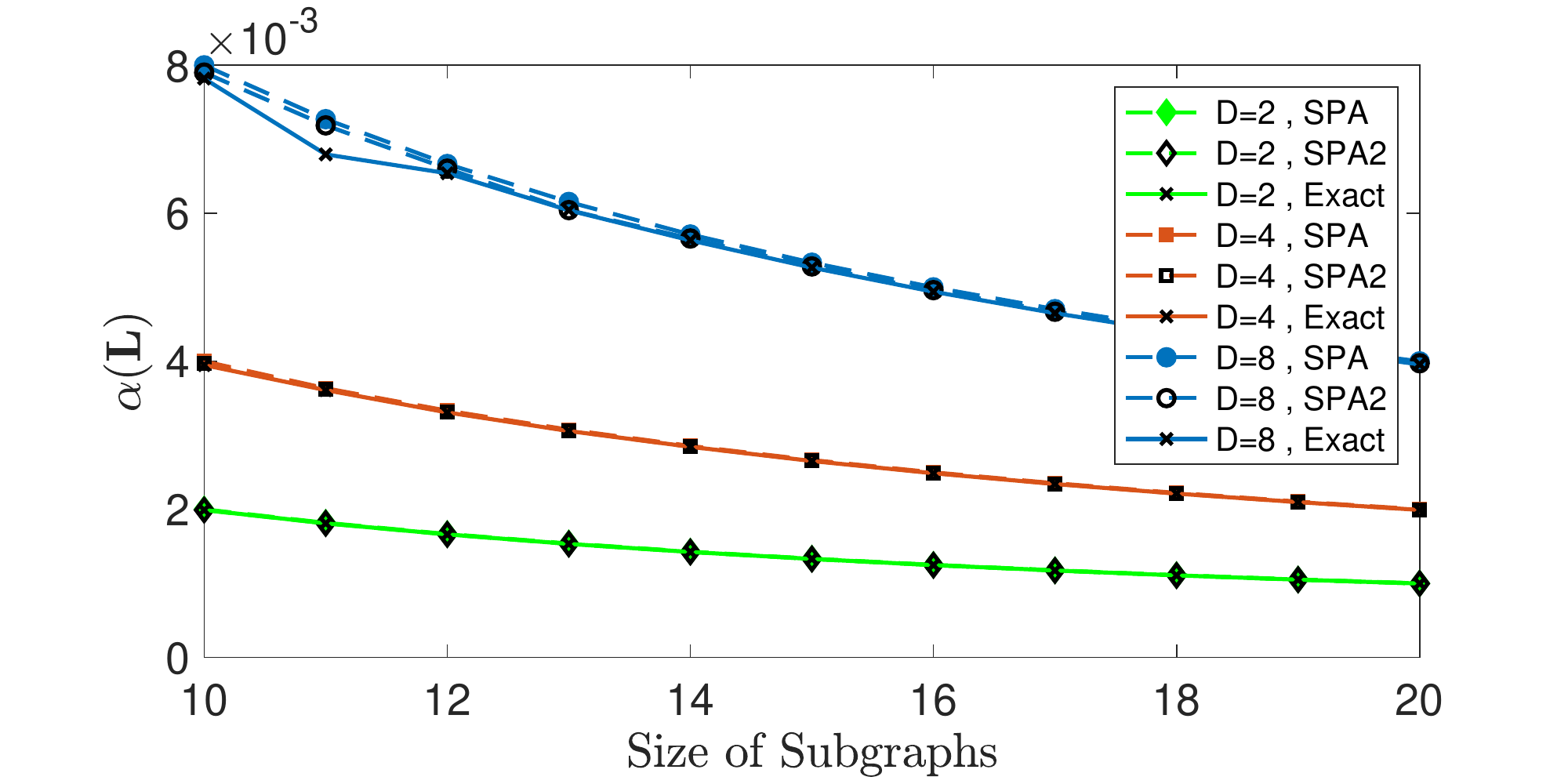}
      \caption{$\epsilon = 0.01$}
      \label{GD1_4}
   \end{subfigure}
      \begin{subfigure}[t]{0.32\linewidth}
      \centering
      \includegraphics[width=\linewidth]{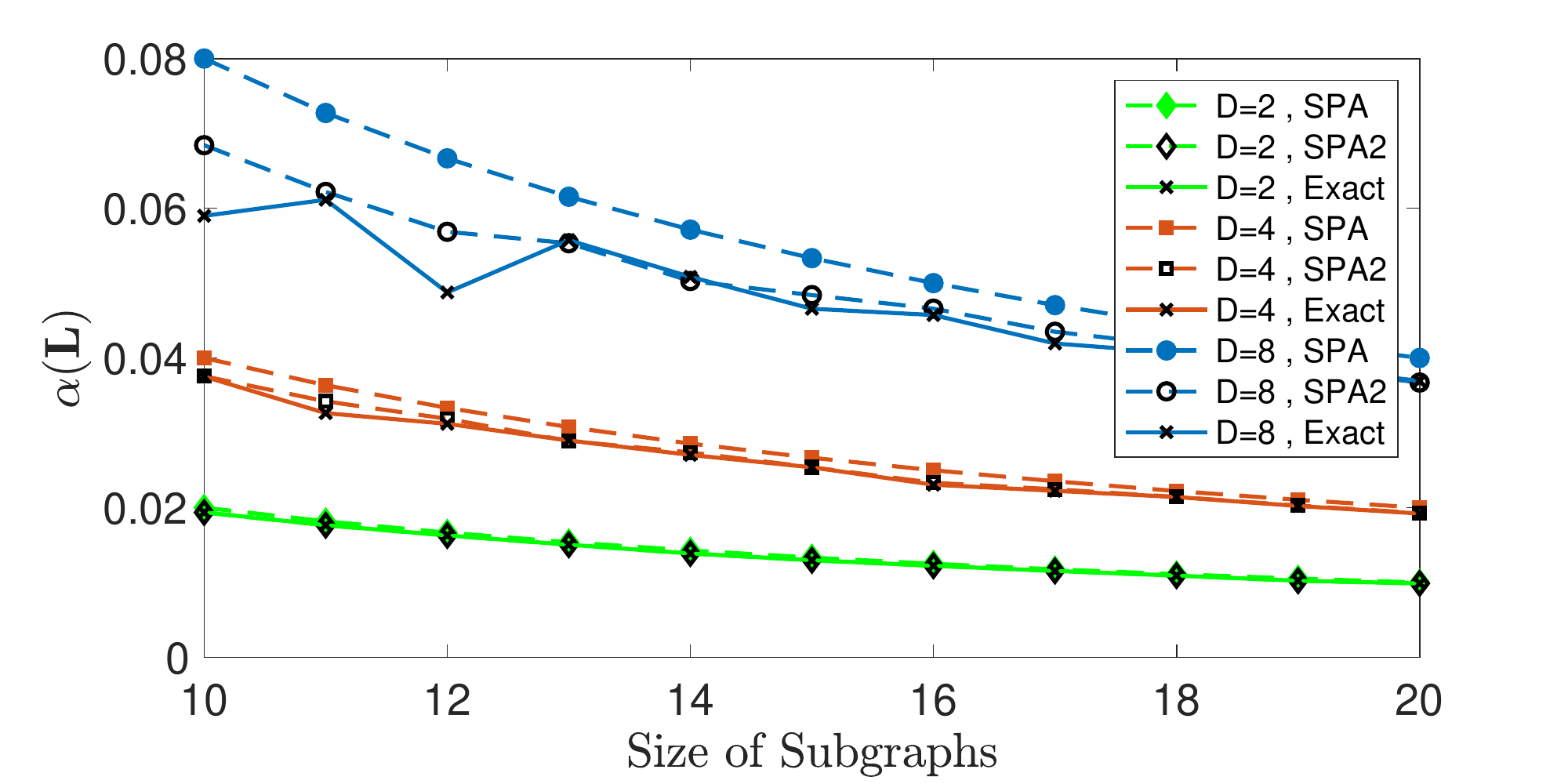}
      \caption{$\epsilon = 0.1$}
      \label{GD1_6}
   \end{subfigure}
   \caption{Spectral gap of the supra-Laplacian matrix, Exact and predicated by perturbation analysis (SPA and SPA2), for composite graphs as the sizes of the individual graphs increase, for various $\epsilon$. The individual  graphs are \Erdos \Renyi random graphs, where an edge exists between each pair of nodes with probability $0.6$, and the connecting graph $\Gcon$ is a complete graph.}
   \label{GD1}
\end{figure*}

\begin{figure*}[htbp]
	\begin{subfigure}[t]{0.32\linewidth} 
	  \centering    
      \includegraphics[width=\linewidth]{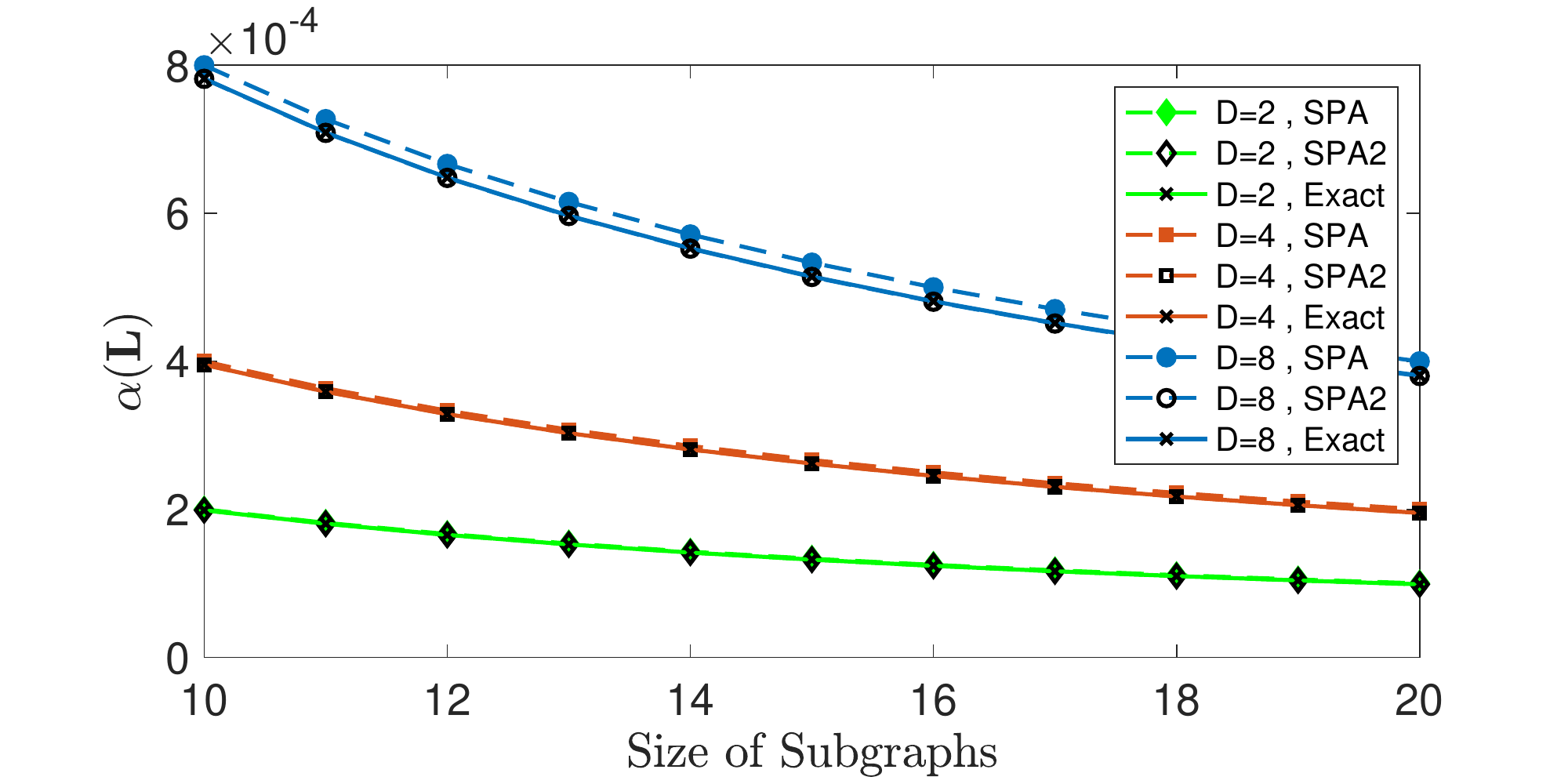}
      \subcaption{$\epsilon = 0.001$}
      \label{GD2_1}
   \end{subfigure}%
   \begin{subfigure}[t]{0.32\linewidth}
      \centering
      \includegraphics[width=\linewidth]{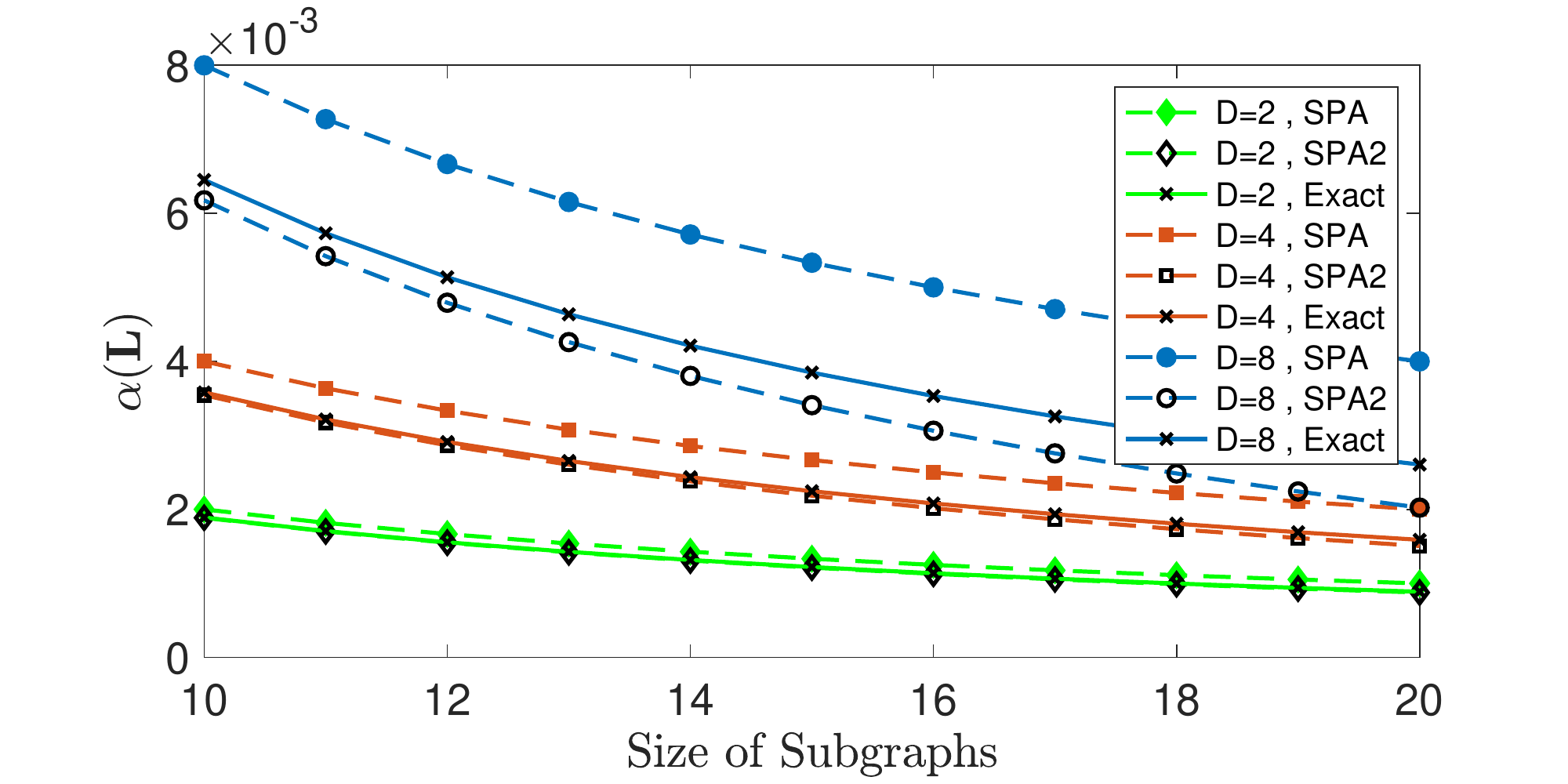}
      \subcaption{$\epsilon = 0.01$}
      \label{GD2_4}
   \end{subfigure}
   \begin{subfigure}[t]{0.32\linewidth}
      \centering
      \includegraphics[width=\linewidth]{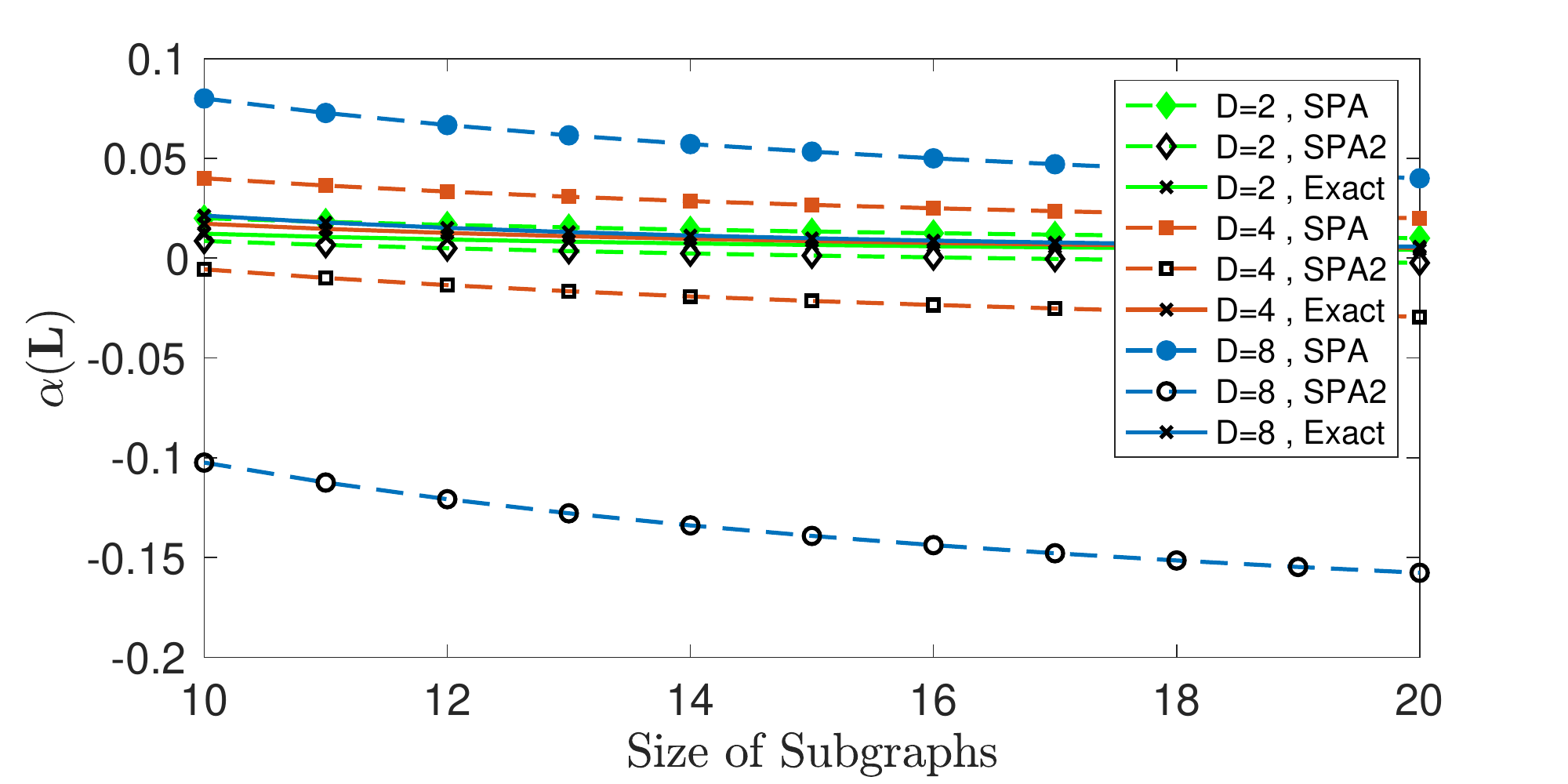}
      \subcaption{$\epsilon = 0.1$}
      \label{GD2_6}
   \end{subfigure}
   \caption{Spectral gap of the supra-Laplacian matrix, Exact and evaluated by perturbation analysis (SPA and SPA2), for composite graphs as the sizes of the individual graphs increase, for various values of $\epsilon$. The individual  graphs are path graphs, and the connecting graph $\Gcon$ is a complete graph.}
   \label{GD2}
\end{figure*}

First, we investigate the spectral gap of the supra-Laplacian matrix in the diffusion dynamics. 
In Fig.~\ref{GD1},  we compare the spectral gap estimated by first order perturbation analysis  (labeled `SPA') and second order perturbation analysis (labeled `SPA2') to the spectral gap directly computed using $\Lc$ (labeled `Exact') for various $\epsilon$.  Each figure shows plots for different numbers of subgraphs, $D=2$, $D=4$, and $D=8$, as the sizes of the subgraphs increase.  Each subgraph is an \Erdos \Renyi random graph with the probability of an edge existing between any two nodes equal to $0.6$.  In each NoN, all subgraphs have the same number of nodes. The connecting graph $\Gcon$ is a complete graph, and the connecting node is chosen uniformly at random in each subgraph.

As expected, the spectral gap decreases as the sizes of the individual subgraphs increase. Also, in general, we see the trend that when $\epsilon$ is held constant, with larger values of $D$, the spectral gap is higher. We explore this phenomenon further in subsequent experiments. 
We observe that the spectral gap generated by first- and second-order perturbation analysis  closely approximates the exact diffusion rate for $\epsilon = 0.001$ to $\epsilon=0.01$. This is in accordance with spectral perturbation theory. The result given by SPA diverges from the exact diffusion rate for a larger value $\epsilon = 0.1$. However, SPA2 still gives good approximation for the spectral gap when $\epsilon = 0.1$.

In Fig.~\ref{GD2}, we show results using the same network scenarios as in Fig.~\ref{GD1}, with the exception that the connecting graphs $\Gcon$ are path graphs. To make the experiment homogeneous, the connecting nodes are selected as end nodes of each path graph.
Again, we note the spectral gap decreases as the size of individual subgraphs increase for all $\epsilon$. The results of SPA and SPA2 closely approximate the exact spectral gap for $\epsilon=0.001$. The result of SPA2 still well approximates the spectral gap for $\epsilon=0.01$, though with less accuracy than in Fig.~\ref{GD1}. 
Both SPA and SPA2 fail to closely approximate the spectral gap for  $\epsilon=0.1$. Thus, we observe that the accuracy of the spectral perturbation analysis depends on the network topology. For each topology, there is some threshold for which, when $\epsilon$ is smaller than this threshold, the approximations are accurate. However, this threshold is different for different NoN topologies.

We also note that, in comparing Fig.~\ref{GD1} and Fig.~\ref{GD2}, it can be observed that the diffusion rate given by SPA coincide for networks of the same size. 
This conforms with our analysis that the first-order approximation of convergence factor of the NoN obtained from spectral perturbation analysis only depends on the sizes of the subgraphs and not on their individual topologies.
 


\begin{figure}[htbp]
      \centering
      \includegraphics[width=.9\linewidth]{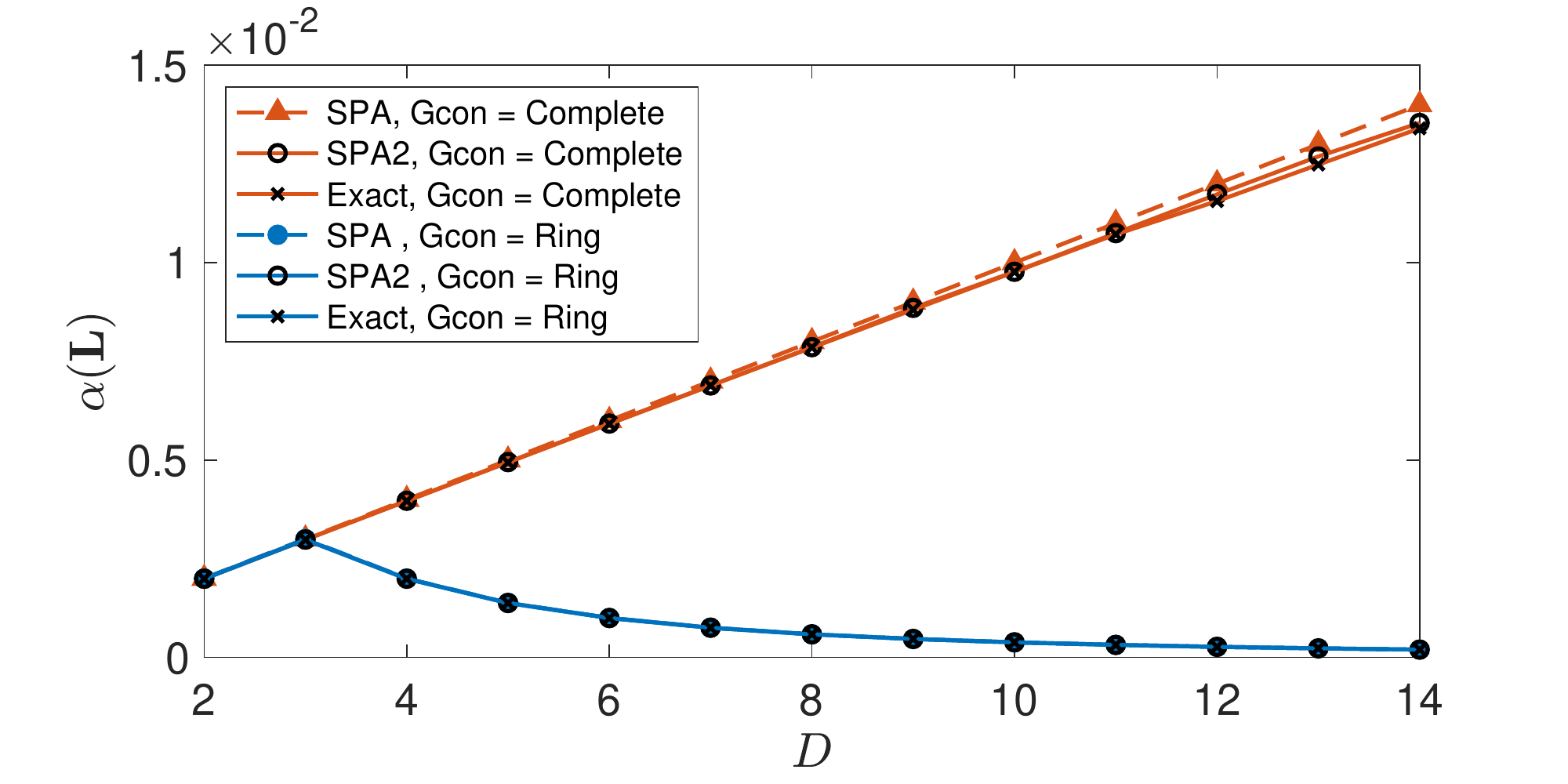}
      \caption{Spectral gap for Exact, SPA, and SPA2, with increasing NoN sizes for ring and complete $\Gcon$ topologies. Subgraphs graphs are \Erdos \Renyi random graphs each with 10 nodes. $\epsilon$ is set to $0.01$.}
      \label{G3}
\end{figure}

In Fig. \ref{G3} we study the dependency of the spectral gap on the topology of $\Gcon$ as the number of subgraphs varies.  Each subgraph is an \Erdos \Renyi random graph with edge probability $0.6$. All subgraphs have $10$ nodes. 
We let $\epsilon= 0.01$, and we compute the convergence factors when $\Gcon$ is complete and when $\Gcon$ is a ring. 

We observe that, when $\Gcon$ is complete, the spectral gap of $\Lc$, in Exact,  SPA , and SPA2, increases with the increase in the number of subgraphs. To better understand this phenomenon, let us assume all subgraphs are of the same size $\Phi$. For $\Gcon$ a complete graph, $\Lcon$ has one eigenvalue of $0$ and $D-1$ eigenvalues equal to $D$. For the SPA diffusion rate given in Theorem~\ref{specGap.thm},  we know that up to first order in $\epsilon$, 
\begin{align}
\rho(\A) & = \epsilon  \left(\frac{D}{N}\right) \lambda_2(\widehat{\Lc}_{con})  = \epsilon  \left(\frac{D}{\Phi}\right).
\end{align}
Since $\Phi$ and $\epsilon$ are held constant, with the increase in $D$, the diffusion rate increases. 
We also note that the diffusion rate when $\Gcon$ is a ring graph is smaller than the diffusion rate when $\Gcon$ is a complete graph.  This can be explained in part by the fact that the algebraic connectivity of a ring graph decreases as its number of nodes increases.

In the following two examples we show that one can use the second order perturbation analysis as a heuristic for choosing connecting nodes to optimize the diffusion rate of the studied diffusion dynamics and the mean square convergence rate of the consensus dynamics.

\begin{figure}[htbp]
      \centering
      \includegraphics[width=.9\linewidth]{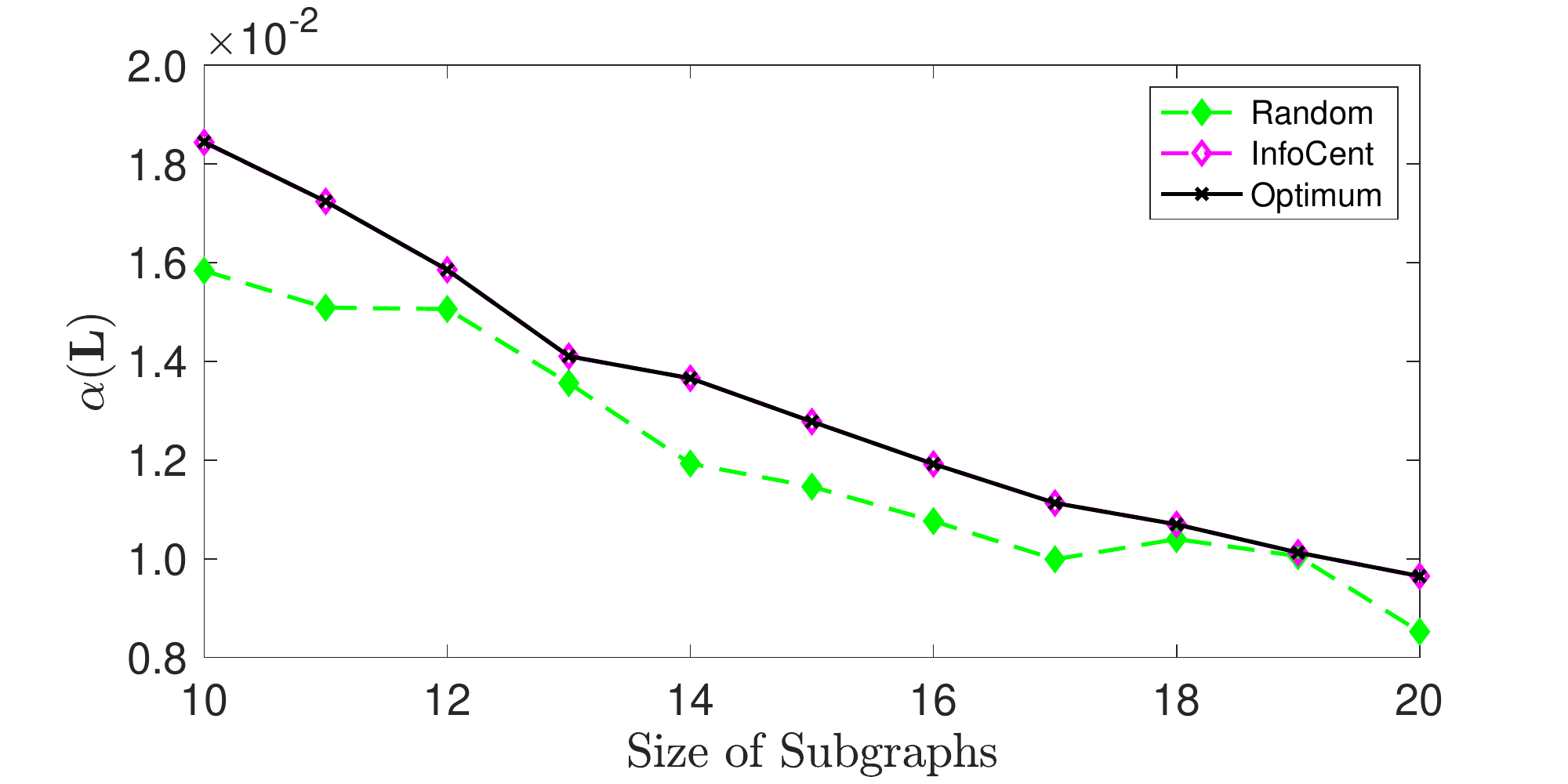}
      \caption{Diffusion rates for the system consists of two subgraphs connected by an edge with bridge nodes selected by different strategy. Subgraphs graphs are \Erdos \Renyi random graphs, where an edge exists between each pair of nodes with probability $0.2$. The diffusion coefficient $\epsilon$ is set to $0.1$.}
      \label{G4}
\end{figure}

\begin{figure}[htbp]
      \centering
      \includegraphics[width=.9\linewidth]{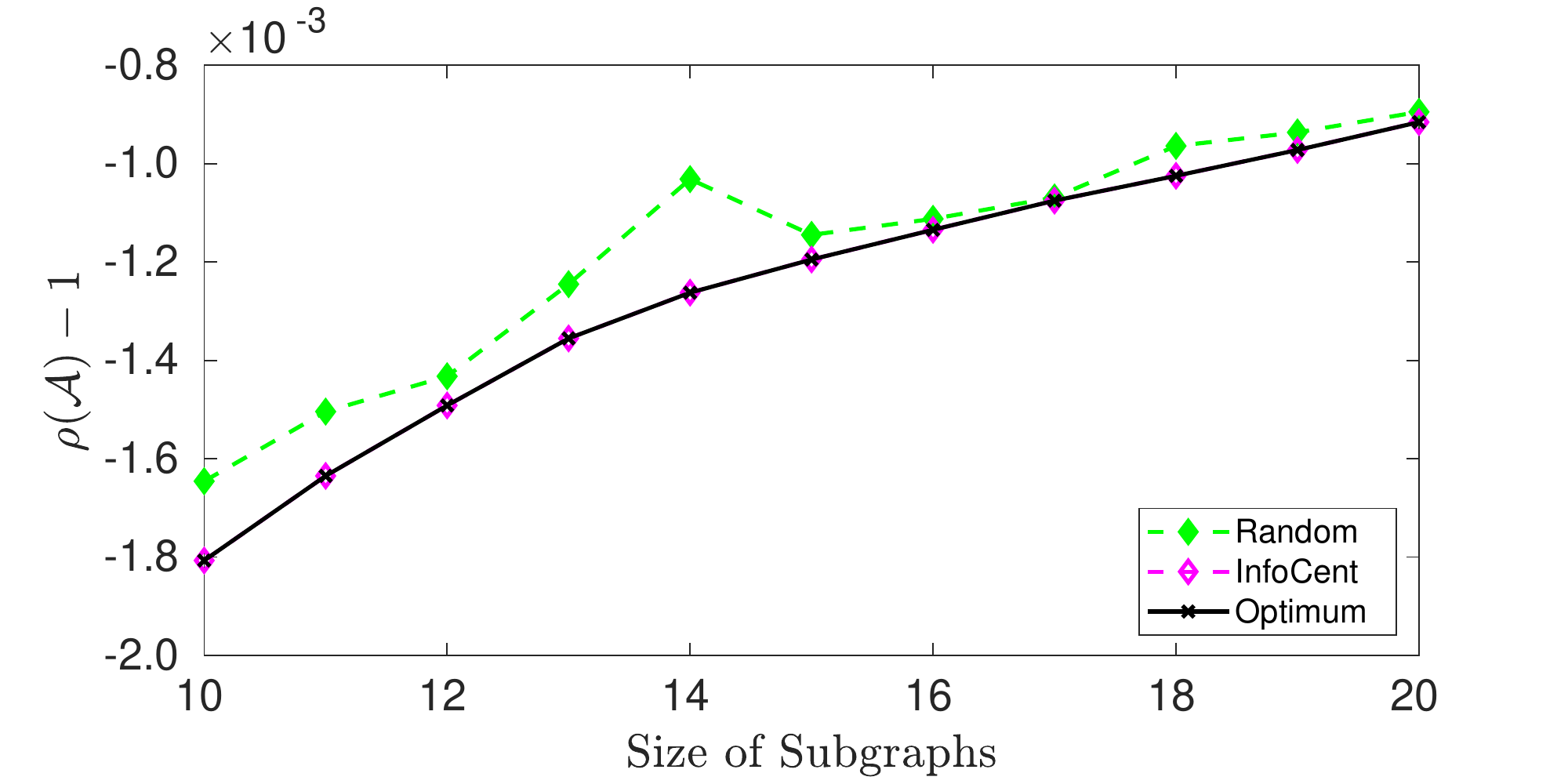}
      \caption{Mean square convergence rates for the system consists of two subgraphs connected by an edge with bridge nodes selected by different strategy. Subgraphs graphs are \Erdos \Renyi random graphs, where an edge exists between each pair of nodes with probability $0.2$. The activation probability of edges in $\Econ$ is $p = 0.1$. $\beta$ takes the value of $\frac{1}{21}$.}
      \label{G5}
\end{figure}

In Fig.~\ref{G4} and Fig.~\ref{G5} we show the exact diffusion rates (given by $\alpha(\Lc)$) and the mean square convergence rates (given by $\rho(\A)$) of systems with different connecting nodes. In both examples we have two \Erdos \Renyi random subgraphs connected by a single edge. The probability that two nodes in the same subgraph are connected is set to $0.2$. And both subgraphs are connected. For the diffusion dynamics, we set $\epsilon = 0.1$. For the consensus dynamics, we let $p=0.1$ and $\beta = \frac{1}{21}$, and $w=\frac{1}{21}$ for all edges in both subgraphs.  In the proposed heuristic, we choose the bridge nodes as the ones with maximum information centrality in each subgraph. We compare the results with the true optimum given by brute-force search, as well as the result of a random choice. The results show that our strategy hits optimal solutions in all occasions, and evidently outperforms the random strategy. We have shown in Theorem \ref{bridgenode.thm} that the second-order approximation of spectral radius of the supra-Laplacian is maximized when connecting nodes are chosen as the ones with largest information centrality. In Fig.~\ref{G4}, we show that by using this result we actually obtain an optimal connecting node in each subgraphs. In Fig.~\ref{G5}, we empirically show that this approach can also be used as a heuristic to find connecting nodes that lead to a good mean square convergence rate.

\section{Conclusion} \label{conclusion.sec}
We have investigated the rate of diffusion in a Network of Networks model, as well as the convergence rate in a consensus NoN with a stochastically switching connecting graph.
Using spectral perturbation analysis, we studied the diffusion rate in a NoN. We showed that the first-order perturbation term is determined by the spectral gap of the generalized Laplacian matrix of the connecting network. In addition, using second-order perturbation analysis, we showed the connection between information centrality and the optimal connecting nodes in subgraphs. 
Finally, we presented numerical results to substantiate our analysis. In future work, we plan to extend our analysis to NoNs with more complex dynamics.



\bibliographystyle{IEEEtran}
\bibliography{references,non_related}

\section{APPENDIX}
\subsection{Analysis of Extended Dynamics} \label{extension.sec}
In Section~\ref{model3.subsubsec}, we assumed that the links in $\Econ$ activate together in a given iteration with probability $p$. 
We now consider a model in which each link is active independently with probability $p$. 
The dynamics of the composite system can then be written as
\begin{align} 
\label{iidactive2.eq}
\xv(\ell+1) = \Ab \xv(\ell) - \sum\limits_{e(r,s) \in \Econ}\delta_{rs}(\ell)\mathbf{B}_{rs}\xv(\ell) 
\end{align}
 where
\begin{align*}
\delta_{rs}(\ell) =
\begin{cases} 1 & \text{ with probability } p \\ 
0 & \text{ with probability } 1-p.
\end{cases}
\end{align*}
Here we let $\delta_{rs}(\ell)$ be mutually independent.

It is straightforward to show that under these dynamics, the autocorrelation matrix $\SSigma$ evolves as 
\[
\SSigma(\ell+1) = \A\left( \SSigma(\ell) \right)
\]
where $\A(X) =\A_0(X) + p \A_1(X)$, with
\begin{align}
\A_0(X) &=  \Pc \Ab \Pc X\Pc \Ab \Pc \nonumber \\
\A_1(X) &= -\sum\limits_{e(r,s) \in \Econ}\mathbf{B}_{rs}X\At  - \At X\sum\limits_{e(r,s) \in \Econ}\mathbf{B}_{rs} \nonumber \\
&~~ + \sum\limits_{e(r,s) \in \Econ}\mathbf{B}_{rs} X \sum \limits_{e(r,s) \in \Econ}\mathbf{B}_{rs}. \nonumber 
\end{align}

We apply spectral perturbation analysis to determine $\rho(\A)$ when $p$ is small.  As before, the spectral radius can be found by examining the perturbations to the 1 eigenvalue of $\A_0$.
These perturbations are given by the spectrum of the matrix $\Fb$. In the case that all graphs have the same number of nodes, $\Fb$ is again a diagonal matrix, with 
\[
f_{ii} = -2 \beta \left( \frac{D}{N}\right) \lambda_i(\widehat{\Lc}_{con})  +  \left(\frac{D}{N}\right)^2 \!\!\left( \sum_{\substack{ (r,s) \in \Econ}}\!\!\!\!\!\! \left( \ub_i^* \Bb_{rs}  \ub_i\right)^2\right). \label{fii-equal.eq}
\]
We note that 
\[
 \sum_{\substack{ (r,s) \in \Econ}}\!\!\!\! \left( \ub_i^* \Bb_{rs}  \ub_i\right)^2 \leq \left(  \sum_{\substack{ (r,s) \in \Econ}} \!\!\!\! \ub_i^* \Bb_{rs}  \ub_i \right)^2  \!\!\!= \! \beta^2 \lambda_i(\widehat{\Lc}_{con})^2.
\]
It follows that an upper bound for $\rho(\A)$, up to first order in $p$, is
\[
\rho(\A) \! \leq \! 1 -p \left(2 \beta  \left(\frac{D}{N}\right) \!\!\lambda_1(\widehat{\Lc}_{con}) \! + \! \beta^2  \left(\frac{D}{N}\right)^2 \!\! (\lambda_1(\widehat{\Lc}_{con}))^2 \!\!\right).
\]
In other words, when $p$ is small, $\rho(\A)$ for a system with the dynamics (\ref{eq:3}) is an upper bound for $\rho(\A)$ for a system with the dynamics (\ref{iidactive2.eq}).

\subsection{Second Order Perturbation Analysis for Mean Square Convergence Rate of the Stochastic Consensus Network}\label{appendix2.sec}
In this appendix we discuss the second order terms in perturbation analysis of the mean-square convergence factor of the system~(\ref{eq:3}).

When $\Fb$ is diagonal, the second order perturbation term associated eigenmatrix $\Wb_i$ could be expressed as 
\begin{align}
C_{i,j}^{(2)} = \sum_{\lambda_j(\A_0) \neq \lambda_i(\A_0)} \frac{\langle \Wb_i,\A_1(\Wb_j)\rangle^2}{\lambda_i(\A_0)-\lambda_j(\A_0)}\,.
\end{align}

\begin{lemma}
The coefficient of second-order perturbation coefficient of eigenvalue \yhy{$\lambda_{ij}(\A)$, $i,j\in\{2,\ldots, D\}$} is
\begin{align}
&C^{(2)}_{ij} = 
(1-\wb_j^*\Bb\wb_j)^2\sum_{\substack{m\in[N]:\\ \lambda_{m}(\At)\neq 1 }}
\frac{1}{\lambda_m(\mathbf{L})}\left(\wb_m^*\Bb\wb_i\right)^2
\nonumber\\
&\qquad + (1-\wb_i^*\Bb\wb_i)^2\sum_{\substack{n\in[N]:\\ \lambda_{n}(\At) \neq 1}}
\frac{1}{\lambda_n(\mathbf{L})}\left( \wb_j^* \Bb \wb_n\right)^2\nonumber\\
&+\sum_{\substack{mn\in[N]:\\ \lambda_{n}(\At)\neq 1,\,\lambda_{m}(\At)\neq 1}}
\frac{\left(\wb_m^*\Bb\wb_i \wb_j^*\Bb \wb_n\right)^2}{\lambda_{m}(\mathbf{L})+\lambda_n(\mathbf{L})-\lambda_{m}(\mathbf{L})\lambda_n(\mathbf{L}) }\,.\label{srpf.eqn}
\end{align}
\end{lemma}

\begin{IEEEproof}
The second order perturbation term of the eigenvalue is attributed to the second order perturbation
terms produce by $\A_1$. Since we have found basis $\wb_i$ such that $\Fb$ is diagonal, we consider the second order term $f^{(2)}_{ij}$
\begin{align}
C^{(2)}_{ij} = & \sum_{\substack{mn:\\ \lambda_{m}(\At)\neq \lambda_{i}(\At)
\\ \lambda_{n}(\At) = \lambda_{j}(\At) }}\frac{\langle\mathbf{W} _{mn}, \A_1(\mathbf{W}_{ij})\rangle^2}{\lambda_{j}(\At)\left(\lambda_{i}(\At)-\lambda_{m}(\At)\right)}\nonumber\\
& + \sum_{\substack{mn:\\ \lambda_{m}(\At) = \lambda_{i}(\At)
\\ \lambda_{n}(\At) \neq \lambda_{j}(\At) }}\frac{\langle\mathbf{W} _{mn}, \A_1(\mathbf{W}_{ij})\rangle^2}{\lambda_{i}(\At)\left(\lambda_{j}(\At)-\lambda_{n}(\At)\right)}\nonumber\\
& + \sum_{\substack{mn:\\ \lambda_{m}(\At)\neq \lambda_{i}(\At)
\\ \lambda_{n}(\At) \neq \lambda_{j}(\At) }}\frac{\langle\mathbf{W} _{mn}, \A_1(\mathbf{W}_{ij})\rangle^2}{\lambda_{i}(\At)\lambda_{j}(\At) - \lambda_{m}(\At)\lambda_{n}(\At)}\,,\label{sop.eqn}
\end{align}
where \yhy{$i,j\in \{2,\ldots, D\}$ corresponds to eigenvalue $1$ of $\At$ with multiplicity $D-1$, and $m,n\in [N]$ corresponds to any eigenvalue of $\At$.}

Since we only consider $C^{(2)}_{ij}$ which is associated with $\lambda_{ij}(\A_0)$ where $\lambda_{i}(\Ab)$ and $\lambda_{j}(\Ab)$ being $1$, and therefore $\lambda_{i}(\Ab)-\lambda_{m}(\Ab)$  and $\lambda_{j}(\Ab)-\lambda_{n}(\Ab)$ are equal to $\lambda_m(\mathbf{L})$ and $\lambda_n(\mathbf{L})$ respectively. We then attain a simplified expression for~(\ref{sop.eqn}),
\begin{align}
f^{(2)}_{ij} = &
\sum_{\substack{mn:\\ \lambda_{m}(\At)\neq \lambda_{i}(\At) \\ \lambda_{n}(\At) = \lambda_{j}(\At)}}
\frac{1}{\lambda_m(\mathbf{L})}\left(-\wb_m^*\Bb\wb_i \left(\wb_j^* \wb_n - \wb_j^*\Bb \wb_n\right)\right)^2\nonumber\\
&
+ \sum_{\substack{mn:\\ \lambda_{m}(\At) = \lambda_{i}(\At)\\\lambda_{n}(\At) \neq \lambda_{j}(\At)}}
\frac{1}{\lambda_n(\mathbf{L})}\left(-(\wb_m^*\wb_i-\wb_m^*\Bb\wb_i) \wb_j^* \Bb \wb_n\right)^2\nonumber\\
&+\sum_{\substack{mn:\\ \lambda_{m}(\At)\neq \lambda_{i}(\At) \\ \lambda_{n}(\At)\neq \lambda_{j}(\At)}}
\frac{\left(\wb_m^*\Bb\wb_i \wb_j^*\Bb \wb_n\right)^2}{\lambda_{m}(\mathbf{L})+\lambda_n(\mathbf{L})-\lambda_{m}(\mathbf{L})\lambda_n(\mathbf{L}) }\nonumber
\end{align}
which can be further written as (\ref{srpf.eqn}).
\end{IEEEproof}

Then we attain the following bounds for $C_{ij}^{(2)}$:
\begin{align}
C^{(2)}_{ij} \geq  &
(1-\wb_j^*\Bb\wb_j)^2\sum_{\substack{m\in[N]:\\ \lambda_{m}(\At)\neq 1 }}
\frac{1}{\lambda_m(\mathbf{L})}\left(\wb_m^*\Bb\wb_i\right)^2
\nonumber\\
&+ (1-\wb_i^*\Bb\wb_i)^2\sum_{\substack{n\in[N]:\\ \lambda_{n}(\At) \neq 1}}
\frac{1}{\lambda_n(\mathbf{L})}\left( \wb_j^* \Bb \wb_n\right)^2\label{sopf_lb.eqn}
\end{align}
and
\begin{align}
& C^{(2)}_{ij} \leq 
(1-\wb_j^*\Bb\wb_j)^2\sum_{\substack{m\in[N]:\\ \lambda_{m}(\At)\neq 1 }}
\frac{1}{\lambda_m(\mathbf{L})}\left(\wb_m^*\Bb\wb_i\right)^2
\nonumber\\
&+ (1-\wb_i^*\Bb\wb_i)^2\sum_{\substack{n\in[N]:\\ \lambda_{n}(\At) \neq 1}}
\frac{1}{\lambda_n(\mathbf{L})}\left( \wb_j^* \Bb \wb_n\right)^2\\
&+\left(\sum_{\substack{m\in[N]:\\\lambda_{m}(\At)\neq 1}}
\frac{1}{\lambda_{m}(\mathbf{L})}
\left(\wb_m^*\Bb\wb_i \right)^2\right)
\left(\sum_{\substack{n\in[N]:\\ \lambda_{n}(\At)\neq 1}}
\left( \wb_j^*\Bb \wb_n\right)^2
\right)\nonumber\,.\label{sopf_ub.eqn}
\end{align}
The upper bound holds since $0<\lambda_m(\mathbf{L})< 1$, and therefore $\lambda_{m}(\mathbf{L}) \leq \lambda_{m}(\mathbf{L})+\lambda_n(\mathbf{L})-\lambda_{m}(\mathbf{L})\lambda_n(\mathbf{L})$.

Then we study the second order refinement of the convergence factor in several classes of NoNs. We omit the proofs of these lemmas due to space limitations.

\begin{corollary}\label{twosubsbound.lemma}
In a composite system consisting of two subgraph and a connecting edge,
With the system dynamics in \yhy{(\ref{eq:3}) satisfying Assumption \ref{activeTogether.assum}}, the second-order perturbation coefficient of the mean square convergence rate of the NoN consensus system,  denoted as $C_{22}^{(2)}$, is bounded by
\begin{align}
C_{22}^{(2)} \geq
&  2\left(1-\beta\left(\theta^{(1)}-\theta^{(2)}\right)^2\right)^2
2 {\beta^2} \nonumber\\
& ~~~~~ \cdot\left(\mathbf{L}_1^{\dag}(s_1,s_1) +
\mathbf{L}_2^{\dag}(s_2,s_2)\right)
\left( \theta^{(1)} - \theta^{(2)}\right)^2\,,\\
C_{22}^{(2)} \leq
&2\left(1-\beta\left(\theta^{(1)}-\theta^{(2)}\right)^2\right)^2
2 {\beta^2}\nonumber\\
&~~~~~\cdot\left(\mathbf{L}_1^{\dag}(s_1,s_1) +
\mathbf{L}_2^{\dag}(s_2,s_2)\right)
\left( \theta^{(1)} - \theta^{(2)}\right)^2\nonumber\\
&+ {\beta^4}\left(\mathbf{L}_1^{\dag}(s_1,s_1) +
\mathbf{L}_2^{\dag}(s_2,s_2)\right)\nonumber\\
&~~~~~\cdot\left(2-\frac{1}{N_1}-\frac{1}{N_2}\right)\left( \theta^{(1)} - \theta^{(2)}\right)^4\,,
\end{align}
where $\mathbf{L}_1^{\dag}(s_1,s_1)$ and $\mathbf{L}_2^{\dag}(s_2,s_2)$ are diagonal entries of the Moore-Penrose inverse of $\mathbf{L}_1$ and $\mathbf{L}_2$. $s_1$ and $s_2$ are the indices of the bridge nodes in each subgraph. In addition, $\theta^{(1)}= \sqrt{\frac{N_2}{N \cdot N_1}}$ and $\theta^{(2)}= -\sqrt{\frac{N_1}{N \cdot N_2}}$.

$\mathbf{L}_1^{\dag}(s_1,s_1)$ and $\mathbf{L}_2^{\dag}(s_2,s_2)$ are minimized when $s_1$ and $s_2$ are chosen as the node with maximum information centrality in each subgraph.
\end{corollary}
\begin{corollary}\label{eqsizebound.lemma}
In  a composite system consisting of $D$ subgraphs $\G_1, \ldots, \G_D$, each with $\frac{N}{D}$ nodes, and a backbone graph $\Gcon$.  With the system dynamics in \yhy{(\ref{eq:3}) satisfying Assumption \ref{activeTogether.assum}}, the second order perturbation coefficient of $\lambda_{ij}(\A)$, $i,j\in\{2,\ldots, D\}$, is bounded by
\begin{align}
& C_{ij}^{(2)} \geq  \beta^2(\lambda_i(\widehat{\Lc}_{con}))^2\left(1-\beta\frac{D}{N}\lambda_j(\widehat{\Lc}_{con})\right)^2\left(\widehat{\wb}_i^*\widehat{\mathcal{I}}\widehat{\wb}_i\right)\nonumber\\
&~~~+ \beta^2(\lambda_j(\widehat{\Lc}_{con}))^2\left(1-\beta\frac{D}{N}\lambda_i(\widehat{\Lc}_{con})\right)^2\left(\widehat{\wb}_j^*\widehat{\mathcal{I}}\widehat{\wb}_j\right) \\
&C_{ij}^{(2)} \leq   \beta^2(\lambda_i(\widehat{\Lc}_{con}))^2\left(1-\beta\frac{D}{N}\lambda_j(\widehat{\Lc}_{con})\right)^2\left(\widehat{\wb}_i^*\widehat{\mathcal{I}}\widehat{\wb}_i\right)\nonumber\\
&~~~+ \beta^2(\lambda_j(\widehat{\Lc}_{con}))^2\left(1-\beta\frac{D}{N}\lambda_i(\widehat{\Lc}_{con})\right)^2\left(\widehat{\wb}_j^*\widehat{\mathcal{I}}\widehat{\wb}_j\right)\nonumber\\
&~~~+ \beta^4 (\lambda_i(\widehat{\Lc}_{con}))^2\left(\frac{D(N-D)}{N^2}\right)\left(\widehat{\wb}_i^*\widehat{\mathcal{I}}\widehat{\wb}_i\right)\,,\label{eqszsndBound}
\end{align}
where the $D\times D$ diagonal matrix $\widehat{\mathcal{I}}$ has its entries $\widehat{\mathcal{I}}(k,k):=\mathbf{L}_k^{\dag}(s_k,s_k)$ for bridge nodes. $s_k$ is the vertex index of the bridge node in subgraph $\G_k$. We note that each $\mathbf{L}_k^{\dag}(s_k,s_k)$ is minimized when the bridge node is chosen as the node with maximum information centrality in that subgraph. Since $\widehat{\mathcal{I}}$ is diagonal, both bounds are minimized when all bridge nodes are chosen with maximum information centrality. 

\end{corollary}

Proposition \ref{twosubsbound.lemma} and \ref{eqsizebound.lemma} show that the second-order perturbation term of mean square convergence rate of system (\ref{eq:3}) is also related to the information centrality of the chosen bridge nodes.

\end{document}